\documentclass[12pt]{article}

\usepackage{epsfig}
\usepackage{amssymb}
\usepackage{graphics}
\let\a=\alpha   \let\b=\beta   \let\g=\gamma   \let\d=\delta
\let\e=\epsilon         \let\q=\theta
      \let\l=\lambda  
           \let\p=\pi      \let\r=\rho
\let\s=\sigma  \let\t=\tau      \let\f=\phi
     \let\y=\psi    
     \let\L=\Lambda

\def\ie{{\it i.e.\ }}

\newcommand{\Zint}{\mathbb{Z}}

\newcommand{\be}{\begin{equation}}
\newcommand{\ee}{\end{equation}}
\newcommand{\bea}{\begin{eqnarray}}
\newcommand{\eea}{\end{eqnarray}}
\newcommand{\ba}{\begin{array}}
\newcommand{\ea}{\end{array}}
\def\nn{\nonumber}
\newcommand{\noN}{\nonumber}

\newcommand{\drawsquare}[2]{\hbox{%
\rule{#2pt}{#1pt}\hskip-#2pt
\rule{#1pt}{#2pt}\hskip-#1pt
\rule[#1pt]{#1pt}{#2pt}}\rule[#1pt]{#2pt}{#2pt}\hskip-#2pt
\rule{#2pt}{#1pt}}
\newcommand{\Yasymm}{\raisebox{-3.5pt}{\drawsquare{6.5}{0.4}}\hskip-6.9pt%
        \raisebox{3pt}{\drawsquare{6.5}{0.4}}}

\newcommand{\YasymmS}{\raisebox{-1.9pt}{\drawsquare{4}{0.4}}\hskip-4.4pt%
        \raisebox{2.1pt}{\drawsquare{4}{0.4}}}
\newcommand{\bYasymmS}{\overline{\YasymmS}}
\begin{document}
%
%
%
%
\begin{titlepage}
\rightline{hep-th/0503044}
%
%
\vskip 2cm
\centerline{{\large\bf A Classification of Toroidal Orientifold
Models}}
\vskip 0.3cm
%
%
\vskip 1cm
\centerline{
P. Anastasopoulos\footnote{panasta@physics.uoc.gr}$^{,\a,\b}$,
A. B. Hammou\footnote{amine@physics.uoc.gr}$^{,\a,\g}$}
\vskip 1cm
\centerline{$^\a$ Department of Physics, University of Crete,}
\centerline{71003 Heraklion, GREECE.}
\vskip  .5cm
\smallskip
\centerline{$^\b$ Laboratoire de Physique Th\'eorique Ecole
Polytechnique,} \centerline{91128 Palaiseau, FRANCE.}
\vskip  .5cm
\smallskip
\centerline{$^\g$
%
%
%
Departement de Physique, Universit\'e des Sciences et Technologie
d'Oran,} \centerline{BP 1505, Oran, El M'Naouer, ALGERIE}
\vskip 1cm
\begin{abstract}

We develop the general tools for model building with orientifolds,
including SS supersymmetry breaking. In this paper, we work out
the general formulae of the tadpole conditions for a class of non
supersymmetric orientifold models of type IIB string theory
compactified on $T^6$, based on the general properties of the
orientifold group elements. By solving the tadpoles we obtain the
general anomaly free massless spectrum.
\end{abstract}
\end{titlepage}

\section{Introduction}

Orientifolds are a generalization of orbifolds \cite{sagn,
Pradisi:1988xd, horava}, where the orbifold symmetry includes
orientation reversal on the worldsheet. The orientifold group
contains elements of two kinds: internal symmetries of the
worldsheet theory forming a group $G_1$ and elements of the form
$\Omega\cdot g$, where $\Omega$ is the worldsheet parity
transformation and $g$ is some symmetry element which belongs to a
group $G_2$. Closure implies that $\Omega\cdot g\cdot\Omega\cdot
g^{\prime} \in G_1$ for $g, g^{\prime} \in G_2$. The full
orientifold group is $G_1+\Omega G_2$. The one loop amplitude
implementing the $\Omega$ projection is interpreted as Klein
Bottle amplitude and has in general ultraviolet divergences
(tadpoles). These ultraviolet divergences are interpreted as
sources in space time, that couple to the massless type IIB
fields, the metric, the dilaton and the R-R forms. They are
localized in sub-manifolds of space-time, known as $orientifold$
$planes$, denoted by $O_p$. These are non dynamical objects
characterized by their charges and tensions. Consistency and
stability of the theory are assured if D-branes are introduced in
a way that guarantees the cancellation of these tadpoles. R-R
tadpole cancellation is equivalent to the vanishing of gauge
charge in a compact space, whereas NS-tadpole cancellation is
equivalent to the vanishing of forces in the D-brane/$O$-plane
vacuum configuration \cite{Kiritsis:2003mc}.


In this work, primarily we study the tadpole conditions for a
class of orientifold groups of the type $G+\Omega ~G$, where $G$
is an orbifold group which contains only geometrical rotation
elements that preserve some supersymmetry.

All the information of the type and the positions of the
orientifold planes are encoded in the Klein bottle amplitudes. We
realize easily that only some specific type of twisted closed
strings couple to the $O$-planes.
In this work, we classify the contribution of the $O$-planes to
the tadpoles by studying the general properties of the elements of
the orbifold group $G$. This classification shows that only
elements $\a\in G$ such that there exist an element $\b\in G$ with
$\a=\b^2$ give non trivial contribution to the tadpoles.

To cancel the aforementioned Klein Bottle UV divergences, we need
to add proper D-branes to the closed string sector (open string
sector). Moreover, the type of D-branes needed in each of the
models together with their contributions to the massless tadpoles
can be similarly classified.

Therefore, we provide general tadpole conditions by studying the
general properties of the elements in $G$. Our general results for
supersymmetric orientifolds agree with the tadpole conditions of
the models already studied in the literature \cite{Gimon:1996ay,
Zwart:1997aj, Aldazabal:1998mr}.

Following the same spirit, we enlarge our study to orientifold
models with spontaneous supersymmetry breaking. For these models,
$G$ contains in addition to rotation elements, also freely acting
ones that break supersymmetry a la Scherk-Schwarz. For simplicity,
the freely acting part we consider is a particular Scherk-Schwarz
deformation by a momentum shift of order two, accompanied by
$(-1)^F$, where $F$ is the spacetime fermion number. We focus on
abelian orientifold groups $G$ where freely and non-freely acting
elements commute. Therefore, since the freely acting elements are
$Z_2$ translations, we restrict our selves for simplicity only to
$Z_2$ rotation factors. Studying the Klein Bottle amplitude in
this case, we realize that in addition to the usual $O$-planes, we
obtain anti-$\bar {O}$-planes when $Z_2$ rotation elements act
longitudinal to the Scherk-Schwarz deformation element. To cancel
all tadpoles we need to add D-branes as well as
$\bar{\textrm{D}}$-antibranes. Finally, we work out the tadpole
conditions based on the general properties of the elements of $G$.
A generalization to other Scherk-Schwarz deformations, as winding
shift or of the type considered in \cite{Scrucca:2001ni}, is not
too difficult to be achieved.

Next step in our study is to solve the tadpole conditions. The
action of the orientifold group on the Chan-Paton factors is made
by the $\g_\a$ matrices. In general, these matrices obey to the
tadpole conditions and to $\g_\a^N=\pm 1$ where $N$ is the
smallest integer for which this equation holds. In general, we can
go to a basis where $\g_\a$ is diagonal with entries the $N$th
roots of unity $\pm 1$. The number of times each entry appears
depends on the tadpole conditions. However, we can give the
general spectra of the models based on the general properties of
the orientifold group. Since the Scherk-Schwarz commutes with the
rotation elements, we can diagonalize the $\g_h$ matrix and study
the way it breaks supersymmetry. Therefore, we provide the general
effect of this element to the supersymmetric representations.



This paper is organized as follows.
In section two we present some generalities about the orientifold
group we are using and discuss the general form of the elements
according to supersymmetry.
In section three we consider in some details the calculations of
the tadpole conditions for the supersymmetric orientifold models.
In section four we discuss the breaking of supersymmetry with a
Scherk-Schwarz deformation.
In section five we give some applications and compare our results
with some of the models existing in the literature.
In section six we discuss the general solutions of the tadpole
conditions and provide the general spectra for supersymmetric and
non-supersymmetric orientifold models.
%
%
%
%
The details of the calculations are presented in the appendices.

\section{The Setup}

Consider the projection of type IIB string theory on $R^4\times T^6$
by the orientifold group $G+\Omega~G$. The orbifold group $G$
contains rotation and translation elements and acts on the
6-dimensional torus $T^6$, which we factorize as $T_1^2\times
T_2^2 \times T_3^2$.
The orbifold action on the complex coordinates $z^i$ of the three
tori with $i=1,2,3$ is given by
\be \a = e^{2\pi i \sum_{i=1}^3 (v_\a^i J^i + R_i~\delta_\a^i
~P^i)}~, \nonumber \ee
with $J^i$ and $P^i$ the generators of rotations and diagonal
translation in each of the internal two torus $T_i^2$ with radius
$R_i$. Let us recall some facts about supersymmetries in
four-dimensional orbifold compactifications \cite{Scrucca:2001ni}.
The basic Majorana-Weyl supercharge $Q$ in $D=10$ fills the ${\bf
16}$ of $SO(9,1)$. This decomposes in $D=4$ into four Majorana
supercharges $Q_n = {Q_n}_L + {Q_n}_R$, transforming each as a
${\bf 2} \oplus {\bf \bar 2}$ under $SO(3,1)$ and together as a
${\bf 4}$ of the maximal $SO(6)$ R-symmetry group. For each
$n=1,2,3,4$, ${Q_n}_{L}$ and ${Q_n}_{R}$ have $SO(6)$ weights
$w_n$ and $-w_n$ respectively, where:
\bea
w_1 = \mbox{$(\frac 12,\frac 12,\frac 12)$} \;,\;\; w_2 =
\mbox{$(\frac 12,-\frac 12,-\frac 12)$} \;,\;\; w_3 =
\mbox{$(-\frac 12,\frac 12,-\frac 12)$} \;,\;\; w_4 =
\mbox{$(-\frac 12,-\frac 12,\frac 12)$} ~.
\eea
A generic orbifold element $\a$ acts as the combination of a
rotation of angle $2 \pi v_i$ and some unspecified shift, in each
of the 3 internal $T^2_i$. Under this action, the four possible
supercharges transform as:
\bea
&&{Q_n}_L \rightarrow  e^{2 \pi i v \cdot w_n} {Q_n}_L ~, \nn \\
&&{Q_n}_R \rightarrow  e^{-2 \pi i v \cdot w_n} {Q_n}_R ~. \eea
Therefore, the supercharge $Q_n$ is left invariant by $\a$ if $v
\cdot w_n$ is an integer, independently of the shift.

For the translation we will consider in this paper only shifts of
order two together with $(-1)^F$, where $F$ is the space-time
fermion number, this element will be denoted through all the paper
by $h$. This is a freely acting orbifold and is a particular
Scherk-Schwarz (SS) deformation, which breaks supersymmetry
spontaneously by giving a mass to the fermions. In general, the
Scherk-Schwarz deformation can be implemented by an order $n$
shift together with a rotation of the same order that breaks
supersymmetry \cite{Scrucca:2001ni, Kiritsis:1997hj} i.e. an
element $v$ such that $\sum_i v_i \neq 0$.

In the direction where a shift acts the only allowed rotations are
those that commute with it. Therefore, in a direction where an
order two shift acts, we will consider at most a rotation by a
$Z_2$ element, denoted by $R_j$ ($R_j:~z^i \to e^{2\pi i
g_i(1-\d_{ij})}z^i$ where $g_i=\pm \frac{1}{2}$. Notice that $R_j$
does not act on the torus $T^2_j$.

Taking into account all these considerations we realize that in
the supersymmetric case, the most general rotation elements $\a$
are such that $v_\a=(v_\a^1,v_\a^2,v_\a^3)$ with say $v_\a^3=0$ or
$v_\a^3\neq 0$.
On the other hand, in the non-supersymmetric cases where we break
supersymmetry by a SS deformation element that acts on the last
torus, the most general rotation elements are of the form
$v_\a=(v_\a^1,v_\a^2,0)$ or $v_\a=(v_\a^1,v_\a^2,\frac{1}{2})$.
The former can be written as $v_\b+g_{i=1,2}$, where $v_\b^3=0$ a
rotation in the $T_1^2\times T_2^2$ torus and $g_{i=1,2}$ a $Z_2$
rotation element. Therefore, without loss of generality we can
take $\a$ such that $v_\a^3=0$.


We would like to study the contribution to the tadpole conditions
of a generic element $\a \in G$. The divergences can be determined
from the vacuum amplitudes on the Klein Bottle (${\cal K}$),
Annulus (${\cal A}$) and M\"obius strip (${\cal M}$). We will
consider two different cases: supersymmetric orientifolds (without
SS $h\notin G$) and non-supersymmetric orientifolds (with SS $h\in
G$).

\section{Supersymmetric Orientifolds}

\subsection*{Klein Bottle}

Let us consider an element $\a$ of $G$. We can work out the
contribution of this element to the Klein Bottle amplitude by
using the trace formula:
\bea {\cal K}_\a = Tr_{U+T}\left[ \Omega \a ~~q^{L_0}
\bar{q}^{\bar{L}_0}\right]~, \label{Klein}\eea
where the subscripts $U$ and $T$ refers to the untwisted and
twisted closed string states of type IIB orbifold model
considered. Due to the presence of $\Omega$ the only states
contributing to ${\cal K}_\a$ are the untwisted states and the
$Z_2$ twisted ones. We remind that by $Z_2$ we mean an order two
rotation elements i.e. $R^2=1$. This element acts always on two
tori since we do not want to break completely supersymmetry. The
twisted states exist only if the orbifold group $G$ contains $Z_2$
factors.

\begin{figure}
\begin{center}
\epsfig{file=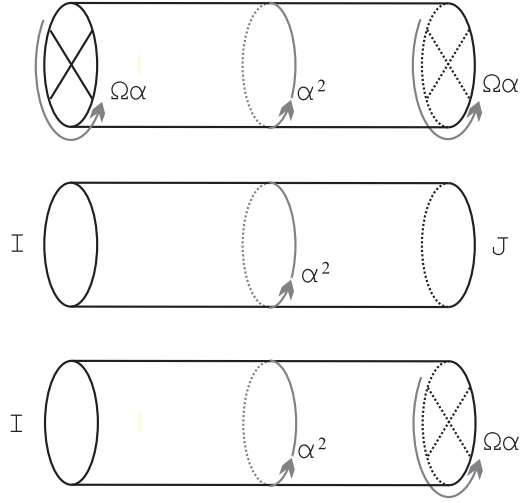,width=70mm}
\end{center}
\caption{Klein-bottle, Annulus and M\"obius strip. The one-loop
amplitudes become tree-level in the transverse picture where an
$\a^2$-twisted closed string propagates between crosscaps and
boundaries.}\label{KleinAnnulusMobius}
\end{figure}

Since the Klein Bottle is equivalent to a cylinder with two
crosscaps (figure.\ref{KleinAnnulusMobius}), then the contribution
of a group element $\a$ corresponds to a propagation of a close
string state projected by $(\Omega \a)^2=\a^2$ for group elements
that commute with $\Omega$ \footnote{ It is because we consider
the orientifold group to be $G+\Omega G$ which implies that
$\Omega$ commutes with all elements of $G$}. If the orbifold group
$G$ contains a $Z_2$ factor denoted by $R$, it produces an extra
contribution since $(\Omega R \a)^2= \a^2$.
%
%
%
To extract the massless tadpole contribution we perform a
modular transformation $l=1/4t$ where $t$ is the loop modulus and
$l$ the cylinder length \cite{Gimon:1996rq, Gimon:1996ay} and then
take the limit $l\to \infty$.
%
%

%
%
%
%
%
%
\begin{itemize}
\item The contribution to the tadpoles from the Klein Bottle
amplitudes of a group element $\a$ such that
$v_\a=(v_\a^1,v_\a^2,0)$ is:
\begin{itemize}
\item[-] When the orbifold group $G$ does not include an $R$ factor,
the only contribution
%
will come from the untwisted sector states:
\bea (1_{NS}-1_{R}) {\cal V}_3 \frac{1}{\prod_{l} 2~\sin 2\pi
v_\a^l} \bigg(\prod_{l} 2~\cos \pi v_\a^l\bigg)^2~, \label{k} \eea
where the factor $\prod_{l} 2~\sin 2\pi v_\a^l$ is related to the
action of $\a$ on Neumann directions \cite{Gimon:1996rq,
Gimon:1996ay} and ${\cal V}_3$ is the regularized volume of the
third torus $T^2_3$.

\item[-] On the other hand, if the group $G$ contains an $R$
factor, we have extra contributions from the $R$-twisted states.
The full contribution for the different cases is:
\begin{itemize}
\item[i.] If there is only one $Z_2$ element that act parallel to
$v_\a$ (only $R_3\in G$):
\bea
(1_{NS}-1_{R}) \frac{1}{\prod_l 2~\sin2\pi v_\a^l} {\cal V}_3
\bigg(\prod_{l} 2~\cos\pi v_\a^l + \prod_{l} 2~\sin\pi
v_\a^l\bigg)^2~. \label{ki} \eea
\item[ii.] If there is only one $R_{i}\in G$ for a given $i=1$ or
$2$ (which $Z_2$ factor acts also on the third torus $T^2_3$):
\bea
(1_{NS}-1_{R})\frac{1}{\prod_{l} 2~\sin2\pi v_\a^l}
\bigg(\sqrt{{\cal V}_3} \prod_{l} 2\cos\pi v_\a^l -
\frac{1}{\sqrt{{\cal V}_3}} 2\cos\pi v_\a^i 2\sin\pi v_\a^j
\bigg)^2~. \label{kii} \eea
\item[iii.] If there are all three possible $Z_2$ factors $R_l\in
G$ with $l=1,2,3$:
\bea
&& (1_{NS}-1_{R})\frac{1}{\prod_l 2\sin2\pi v_\a^l}
\Bigg\{\sqrt{{\cal V}_3} \bigg(\prod_{l} 2\cos\pi v_\a^l +
\prod_{l} 2\sin\pi v_\a^l\bigg)
\nn\\
&&\qquad \qquad ~~~ - \frac{1}{\sqrt{{\cal V}_3}}\sum_{i\neq
j=1,2} \epsilon_{ij}~2\cos\pi v_\a^i 2\sin\pi v_\a^j \Bigg\}^2~.
\label{kiii} \eea
where $\epsilon_{12}=-\epsilon_{21}=1$.
\end{itemize}
\end{itemize}
%
%
%
All the amplitudes above are proportional to $(1_{NS}-1_{R})$ and
their multiplicatives appear as perfect squares
\cite{Pradisi:1988xd, Angelantonj:2002ct}. We should mention that
for this kind of orbifold action all the amplitudes are volume
dependant (${\cal V}_3$).
They are of the general form:
\be (1_{NS}-1_{R}) \left[ K_1 \sqrt{{\cal V}_3}+{K_2 \over
\sqrt{{\cal V}_3}}\right]^2 ~,\label{KB-general}\ee
where $K_1$ and $K_2$ are constants encoding the information about the
orbifold projection.
\item Consider now an element that acts on all the three tori
$v_\a=(v_\a^1,v_\a^2,v_\a^3)$ where $v_\a^{l=1,2,3}\neq 0$ or
$1/2$. The contribution to the tadpoles depends again on the
existence of an $R$ factor in $G$:
%
%
%
%
\begin{itemize}
\item[-] If $G$ contains no $R$ factors:
\bea (1_{NS}-1_{R}) \frac{1}{\prod_l 2~\sin 2\pi v_\a^l}
\bigg(\prod_l 2~\cos \pi v_\a^l\bigg)^2~. \label{kk} \eea
\item[-] If $G$ contains $R$ factors, then:
\begin{itemize}
\item[i.] if it contains only one $Z_2$ factor $R_i\in G$ for a
given $i$:
\bea (1_{NS}-1_{R}) \frac{1}{\prod_l 2~\sin2\pi v_\a^l}
\bigg(\prod_l 2~\cos\pi v_\a^l + 2~\cos\pi v_\a^i \prod_{l\neq i}
2~\sin\pi v_\a^l\bigg)^2~. \label{kki} \eea
\item[ii.] If $G$ contains all three possible $Z_2$ factors
$R_l\in G$ for $l=1,2,3$:
\bea (1_{NS}-1_{R}) \frac{1}{\prod_l 2\sin2\pi v_\a^l}
\bigg(\prod_l 2\cos\pi v_\a^l + \sum_i 2\cos\pi v_\a^i
\prod_{l\neq i} 2\sin\pi v_\a^l\bigg)^2~. \label{kkii} \eea
\end{itemize}
\end{itemize}
All the amplitudes are again perfect squares as they should
\cite{sagn, Pradisi:1988xd, Angelantonj:2002ct}.
\end{itemize}
\begin{figure}
\begin{center}
\epsfig{file=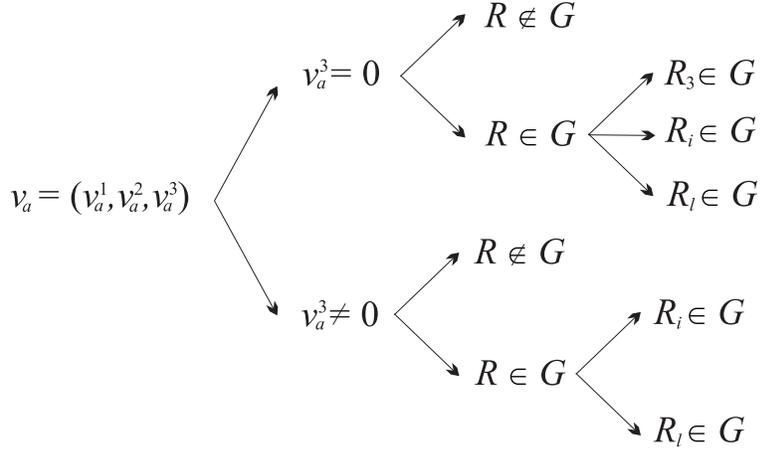,width=100mm}
\end{center}
\caption{We classify the tadpole conditions according to the kind
of elements belonging to the orbifold group $G$. $i=1$ or $2$ and
$l=1,2,3$.}\label{Classification}
\end{figure}

In orientifold models, tadpoles arise as divergences at the
one-loop level, which are interpreted as inconsistencies in the
field equations for the R-R potentials in the theory. These
tadpoles can be regarded as the emission of an R-R closed string
state from a D$p$-brane (disc), a source of (p+1)-form R-R
potential and from an orientifold plane (${\bf RP}^2$), which
carries R-R charges. The contributions we have obtained above come
from $Op$-planes sources of twisted (p+1)-form R-R potentials. For
non trivial twists, the string zero modes vanish, and therefore
the D$p$-brane is forced to sit at the fixed point. It simply
means that the twisted R-R potential can not propagate in
spacetime. Depending on the tension and charge, there are four
kinds of orientifold planes: $O^+$ ($O^-$) with negative
(positive) tension and charge and $\bar{O}^+$ ($\bar{O}^-$) with
negative (positive) tension and positive (negative) charge
\cite{Angelantonj:2002ct}.
In this paper we will meet three kinds of these
orientifold-planes, namely $O^-$, $\bar{O}^-$ and $O^+$. The later
appear in models with broken supersymmetry by an order two
Scherk-Schwarz deformation \cite{Angelantonj:2002ct,
Scrucca:2001ni, Antoniadis:1998ki}.
The $\bar{O}^+$ orientifold planes appear in orientifold models of
type IIB string theory with a discrete NS antisymmetric tensor
switched on (discreet torsion) \cite{Antoniadis:1998ki}.

\subsection*{Annulus}

To cancel the aforementioned Klein Bottle UV divergences
-tadpoles- we need to add D-branes to the spectrum (open string
sector). These are a bunch of D9-branes and D5$_i$-branes extended
along the $T_i^2$ torus and siting on the $R_i$-fixed points when
the group $G$ contains $R_i$-factors \footnote{We should mention
that D7 and D3 branes are obtained in orientifold models where
$\Omega$ acts together with a reflection  element $I_2$ that gives
a minus sign to the coordinates of one of the $T_i^2$ torus.}.
The Annulus amplitudes can be easily computed for all kinds of
D-branes existing in the theory, with the contribution of the
group element $\a$ given by the trace formula:
\bea {\cal A}_{IJ,\a} = Tr_{IJ} \left[\a ~q^{L_0} \right]~, \eea
where now the trace is over all open string states attached
between $I$ and $J$ D-branes for $I,J=9,5_i$ siting at $\a$-fixed
point.
%
%
%
To extract the tadpole contributions we need to perform a modular
transformation to the transverse channel $l=1/2t$ and then take
the limit $l\to \infty$ \cite{Gimon:1996rq}.

Similar to the Klein-bottle cases, we classify the contributions
to the tadpoles by the existence of $Z_2$ rotation elements in $G$
(figure.2):
\begin{itemize}
\item For an element $\a$ such that: $v_\a=(v_\a^1,v_\a^2,0)$ we
have the following contribution to the tadpoles:
\begin{itemize}
\item[-] if $G$ contains no $Z_2$-factors, the only contribution
to the annulus amplitude is coming from the 99 strings.
\bea (1_{NS}-1_R){\cal V}_3 \frac{1}{\prod_{l} 2~\sin \pi v_\a^l}
Tr[\g_{\a,9}]^2~. \label{a} \eea
\item[-] if $G$ contains $R$-factors, then we have also
contributions from the corresponding D5-branes. As in the Klein
Bottle case, we have the following cases:
\begin{itemize}
\item[i.] Only $R_3\in G$:
\bea (1_{NS}-1_R){\cal V}_3 \frac{1}{\prod_{l} 2~\sin \pi
v_\a^l}\Big(Tr[\gamma_{\a,9}]+ \prod_{l} 2\sin\pi
v_\a^l~Tr[\gamma_{\a,5_{3}}]\Big)^2~. \label{ai} \eea
\item[ii.] Only $R_i\in G$ for a given $i=1$ or $2$:
\bea (1_{NS}-1_R) \frac{1}{\prod_{l} 2~\sin \pi
v_\a^l}\Big(\sqrt{{\cal V}_3}Tr[\gamma_{\a,9}]-
\frac{1}{\sqrt{{\cal V}_3}}2\sin\pi
v_\a^j~Tr[\gamma_{\a,5_{i}}]\Big)^2~. \label{aii} \eea
\item[iii.] All three $Z_2$ factors $R_l\in G$ with $l=1,2,3$:
\bea
&& (1_{NS}-1_{R}) \frac{1}{\prod_l 2\sin2\pi v_\a^l}
\Bigg[\sqrt{{\cal V}_3} \Big(Tr[\gamma_{\a,9}]+ \prod_{l} 2\sin\pi
v_\a^l Tr[\gamma_{\a,5_3}]\Big)
\nn\\
&&~~~~~~~~~~~~~~~~~~~~~~~~~~~~~~ - \frac{1}{\sqrt{{\cal
V}_3}}\sum_{i\neq j=1,2} 2\sin\pi v_\a^j
Tr[\gamma_{\a,5_i}]\Bigg]^2~. \label{aiii} \eea
\end{itemize}
\end{itemize}
The amplitudes are again proportional to zero: $(1_{NS}-1_{R})$,
reflecting the fact that the orientifold group action preserve
supersymmetry. The multiplicative factor is a function of the
volume of the unaffected torus:
\be (1_{NS}-1_{R}) \left[ A_1 \sqrt{\cal V}_3+{A_2 \over
\sqrt{\cal V}_3}\right]^2~. \label{A-general}\ee
$A_1$ and $A_2$ are functions of the traces of the Chan-Paton
factors $Tr[\g_{\a, I}]$. $A_1$ is the contribution of the
Newmann-Newmann strings longitudinal to the unaffected torus (they
have NN boundary conditions in this torus), and is proportional to
$Tr[\g_{\a,9}]$ and $Tr[\g_{\a,5_3}]$. Whereas, $A_2$ is the
contribution of the Dirichlet-Dirichlet strings transverse to
${\cal V}_3$, and is function of $Tr[\g_{\a,5_i}]$ for $i=1,2$.

\item For an element $\a$ such that $v_\a=(v_\a^1,v_\a^2,v_\a^3)$
the contribution to the tadpoles from the Annulus amplitudes are
as follows:

%
%
\begin{itemize}
\item[-] when $G$ contains no $Z_2$ factors, we have just the
contribution of the 99 strings.
\bea (1_{NS}-1_R) \frac{1}{\prod_{l} 2~\sin \pi v_\a^l}
Tr[\gamma_{\a,9}]^2~. \label{aa} \eea
\item[-] when $G$ contains $R$ factors, then:
\begin{itemize}
\item[i.] If there is only one $R_i\in G$ (for a given $i$), we
have:
\bea (1_{NS}-1_R) \frac{1}{\prod_{l} 2~\sin \pi
v_\a^l}\Big(Tr[\gamma_{\a,9}]+ \prod_{l\neq i} 2\sin\pi
v_\a^l~Tr[\gamma_{\a,5_{i}}]\Big)^2~. \label{aai} \eea
\item [ii.] If all three $R_l\in G$ (with $l=1,2,3$), we should
include its corresponding D5$_l$-branes:
\bea (1_{NS}-1_{R}) \frac{1}{\prod_{l} 2\sin2\pi v_\a^l}
\Big(Tr[\gamma_{\a,9}]+ \sum_{i=1}^3 \prod_{l\neq i} 2\sin\pi
v_\a^l Tr[\gamma_{\a,5_i}]\Big)^2~. \label{aaii} \eea
\end{itemize}
\end{itemize}
The structure of these amplitudes is similar to (\ref{A-general})
without the volume dependance and with the product extended over
$l=1,2,3$.
\end{itemize}

\subsection*{M\"obius Strip}

Finally, the contribution of the group element $\a$ to the
M\"obius strip amplitude can be computed from the trace formula:
\bea {\cal M}_{I,\a} = Tr_I \left[\Omega \a ~q^{L_0}\right]~, \eea
where the trace is over open strings attached on the D$I$-brane.
To extract the contribution to the tadpoles we must perform a
modular transformation to the transverse channel by $P=T S T^2 S
T$ where $T:\t \to \t+1$ and $S:\t \to - 1/\t$, $l=1/8t$. Finally,
we take the UV limit $l \to\infty$.
%
%

The M\"obius strip transverse channel amplitude is also the mean
value of the transverse channel Klein Bottle and Annulus
amplitudes \cite{Pradisi:1988xd, Angelantonj:2002ct}. Therefore,
extracting the UV limit in the M\"obius strip amplitude and
comparing with the Klein Bottle and Annulus amplitudes, we obtain
constraints on the matrices $\g_{\a,I}$ and $\g_{\Omega.\a,I}$:
\bea &&Tr[\gamma^T_{\Omega \a,9}\gamma^{-1}_{\Omega
\a,9}]=Tr[\gamma_{\a^2,9}]~,
\nn\\
&&Tr[\gamma^T_{\Omega R_i \a,9}\gamma^{-1}_{\Omega R_i \a,9}]=
-Tr[\gamma_{\a^2,9}]~,
\nn\\
&&Tr[\gamma^T_{\Omega \a,5_i}\gamma^{-1}_{\Omega \a,5_i}]=
-Tr[\gamma_{\a^2,5_i}]~,
\nn\\
&&Tr[\gamma^T_{\Omega R_i \a,5_i}\gamma^{-1}_{\Omega R_i \a,5_i}]=
Tr[\gamma_{\a^2,5_i}]~,
\nn\\
&&Tr[\gamma^T_{\Omega R_j \a,5_i}\gamma^{-1}_{\Omega R_j \a,5_i}]=
-Tr[\gamma_{\a^2,5_i}]~, \label{consta} \eea
where in the last equation $i\neq j$ and $i,j= 1,2,3$.
These constraints appears to be the same for both
$v_\a=(v_\a^1,v_\a^2,0)$ and $v_\a=(v_\a^1,v_\a^2,v_\a^3)$.
Details of the calculations are reported in the appendix.

\subsection*{Tadpole conditions:}\label{Tadpoles-NoSS}

The massless part of the transverse channel amplitudes
$\tilde{\cal K}_\a +\tilde{\cal A}_\a +\tilde{\cal M}_\a$ gives
the tadpole conditions.
The tadpole conditions for a given element $\a^2$ are:
\begin{itemize}
\item If $\a$ is such that $v_\a=(v_\a^1,v_\a^2,0)$, then:
\begin{itemize}
\item[-] when $G$ contains no $Z_2$ factors:
\bea Tr[\gamma_{\a^2,9}]= 32 \prod_l \cos \pi v_\a^l~.
\label{tvk1} \eea
\item[-] when $G$ contains $Z_2$ factors, then we have the
following cases:
\begin{itemize}
\item[i.] If only $R_3\in G$:
\bea Tr[\g_{\a^2,9}]+ 4\prod_{l} \sin2\pi
v_\a^l~Tr[\g_{\a^2,5_{3}}]= 32~ \bigg(\prod_{l} \cos\pi v_\a^l
+\prod_{l} \sin\pi v_\a^l\bigg)~. 
%
\label{tvk2i} \eea
\item[ii.] If only $R_{i}\in G$ for a given $i=1$ or $2$:
\bea
Tr[\gamma_{\a^2,9}]&=& 32~\prod_{l} \cos\pi v_\a^l~ ,\nn\\
2\sin2\pi v_\a^j~Tr[\gamma_{\a^2,5_{i}}]&=& 32~\cos\pi v_\a^i
\sin\pi v_\a^j~. \label{tvk2ii} \eea
\item[iii.] If all three $R_l\in G$ with $l=1,2,3$:
\bea Tr[\gamma_{\a^2,9}]+4 \prod_{l} \sin2\pi
v_\a^lTr[\gamma_{\a^2,5_3}] &=& 32~\bigg(\prod_{l} \cos\pi v_\a^l
+ \prod_{l} \sin\pi v_\a^l\bigg)~ ,
\nn\\
\sum_{i\neq j =1,2} 2\sin2\pi v_\a^j Tr[\gamma_{\a^2,5_i}]&=&
32~\sum_{i\neq j =1,2} \epsilon_{ij} \cos\pi v_\a^i \sin\pi
v_\a^j~.
\label{tvk2iii} \eea
\end{itemize}
\end{itemize}
\item If $\a$ is such that $v_\a=(v_\a^1,v_\a^2, v_\a^3)$ then:
\begin{itemize}
\item[-] when $G$ does not contain any $Z_2$ factors:
\bea Tr[\g_{\a^2,9}]= 32~\prod_l \cos \pi v_\a^l~, \label{tvl1}
\eea
\item[-] when $G$ does contain $Z_2$ factors, then:
\begin{itemize}
\item[i.] $R_i\in G$ for a given $i$:
\bea Tr[\gamma_{\a^2,9}]&+& 4\prod_{l\neq i} \sin2\pi
v_\a^l~Tr[\gamma_{\a^2,5_{i}}]\nn\\
&=& 32~ \bigg(\prod_l \cos\pi v_\a^l + \cos \pi v_\a^i
\prod_{l\neq i} \sin\pi v_\a^l\bigg)~.
\label{tvl2i}\eea
\item[ii.] $R_l\in G$ with $l=1,2,3$:
\bea Tr[\gamma_{\a^2,9}]&+& 4\sum_{i=1}^3 \prod_{l\neq i} \sin2\pi
v_\a^l Tr[\gamma_{\a^2,5_i}]\nn\\
&=& 32~ \bigg(\prod_l \cos\pi v_\a^l + \sum_i \cos\pi v_\a^i
\prod_{l\neq i} \sin\pi v_\a^l\bigg)~.
\label{tvl2ii} \eea
\end{itemize}
\end{itemize}
\end{itemize}
Note that all the tadpole conditions hold for both NS and R
sectors according to supersymmetry.

The tadpole condition for group elements that are not square of
some other element of $G$ (there is no element $\b \in G$ such
that $\a = \b^2$), will receive contribution only from the Annulus
amplitude. When this element is such that $v=(v_\a^1,v_\a^2,0)$ or
$v=g_3+v_\a$, the tadpole conditions will be the same as before
(\ref{tvk1}-\ref{tvk2iii}), with zero on the right hand side. For
elements such that $v=g_i+v_\a$ it is not difficult to work out
the tadpole conditions, leading to:
\bea Tr[\gamma_{R_i\a,9}]&+& 4\sin\p v_\a^i \cos\p v_\a^j
Tr[\gamma_{R_i\a,5_3}]
\nonumber\\
&+&2\cos\pi v_\a^j Tr[\gamma_{R_i\a,5_i}]+2\sin\p v_\a^i
Tr[\gamma_{R_i\a,5_j}]=0~, \label{tgiv} \eea
where $i\neq j =1,2$ and the different terms exist only if the
corresponding $R_i$ factor does.
If $v_\a=(v_\a^1,v_\a^2,v_\a^3)$, the tadpole conditions are the
same as (\ref{tvl1}-\ref{tvl2ii}) without the right hand side (\ie
the right hand side is zero).

In Section \ref{Examples} we will give some applications of the
formulae we have obtained in this section and compare with the
supersymmetric orientifolds studied in the literature.

\section{Breaking Supersymmetry with momentum shifts}
\label{MomentaShift}

In this section we include the Scherk-Schwarz (SS) deformation
which is a translation in one direction in the torus $T^2_3$ by a
momentum shift of order two accompanied by $(-1)^F$ where $F$ is
the spacetime fermion number. This deformation is compatible only
with an orbifold action that commutes with it, therefore, we will
restrict ourselves to elements of the form $v=(v_1,v_2,0)$ or $v+
g_i$ where $g_i$ is the shift vector of a $Z_2$ element that
leaves the coordinates of $T^2_i$ invariant and gives a minus sign
to the other compact coordinates.

\subsection*{Klein Bottle}\label{KleinBottle+SS}

Since $\Omega$ exchange the left and the right moving modes, it
acts on the momentum and windings as:
\bea \Omega~|m,n\rangle \rightarrow ~|m,-n\rangle~. \eea
Therefore, the invariant states are those with vanishing winding
modes i.e. $n=0$. The shift acts on the zero modes as $(-1)^m
\Lambda_{m,n}$ where $\Lambda_{m,n}$ is the Narain lattice sum. It
is easy to see that the $h$ twisted sector with the zero mode part
being $\Lambda_{m,n+\frac{1}{2}}$ does not survive the $\Omega$
projection. However, if $\Omega$ is acting together with a $Z_2$
element, the combined action gives:
\bea \Omega R~|m,n\rangle \rightarrow ~|-m,n\rangle ~.\eea
The invariant states in this case are those with vanishing
momentum i.e. $m=0$. Therefore, the $h$ twisted sector survives
this projection if $h$ and $R$ acts in the same direction.

It is not difficult to see that the tadpoles in this model
correspond to $O5$ orientifold planes sitting on the $R$-fixed
points (as in the previous section) and $\bar{O}5$ orientifold
planes sitting on the $Rh$-fixed points \cite{Scrucca:2001ni,
Antoniadis:1998ki}.
To extract the massless tadpole contribution we need to perform a
modular transformation $l=1/4t$ and then take $l\to \infty$.
Except from (\ref{k}-\ref{kiii}) we will also have contributions
from the $h$ twisted sector
%
%
%
as well as $R_i h$, they both contribute $(1_{NS}+1_R)$ to the
tadpoles reflecting the breaking of supersymmetry.
%
%
%
\begin{itemize}
\item[-] When the orbifold group $G$ does not contain a $Z_2$
element, the contribution to the Klein Bottle will result only
from the untwisted sector as for the case without SS. There is no
contribution from states ${\cal K}_{h\a}$ (\ref{Klein}) due to the
shift (it gives rise only to massive states). Therefore, the
contribution is the same as the one obtained in the supersymmetric
case (\ref{k}).
%
%
\item[-] when $G$ contains an $R$ factor, we have also
contributions from the $R$ twisted states:
\begin{itemize}
\item[i.] If only $R_3\in G$,
%
%
the contribution is exactly as before (\ref{ki}) (without SS),
because the Scherk-Schwarz deformation is acting transverse to
$R_3$. The $R_3 h$ twisted states do not contribute since they are
massive.
\item[ii.] If only $R_{i}\in G$ for a given $i=1$ or $2$:
\bea &&\frac{1}{\prod_{l} 2~\sin2\pi v_\a^l} \bigg[1_{NS}
\bigg(\sqrt{{\cal V}_3} \prod_{l} 2\cos\pi v_\a^l
-\frac{2}{\sqrt{{\cal V}_3}}2\cos\pi v_\a^i 2\sin\pi
v_\a^j\bigg)^2
\nn\\
&&~~~~~~~~~~~~~~~~~-1_R {\cal V}_3 \bigg(\prod_{l} 2~\cos\pi
v_\a^l\bigg)^2\bigg]~ , \label{khii} \eea
\item[iii.] If all three $R_l\in G$ with $l=1,2,3$:
\bea
&&\frac{1}{\prod_{l} 2\sin2\pi v_\a^l} \bigg\{1_{NS}
\bigg[\sqrt{{\cal V}_3} \bigg(\prod_{l} 2\cos\pi v_\a^l +
\prod_{l} 2\sin\pi v_\a^l\bigg)
\nn\\
&&- \frac{2}{\sqrt{{\cal V}_3}}\sum_{i\neq j =1,2}
\epsilon_{ij}~2\cos\pi v_\a^i 2\sin\pi v_\a^j \bigg]^2-1_R {\cal
V}_3 \bigg(\prod_{l} 2~\cos\pi v_\a^l +\prod_{l} 2~\sin\pi
v_\a^l\bigg)^2\bigg]\bigg\}~. \nn\\
\label{khiii} \eea
\end{itemize}
\end{itemize}
All the amplitudes are perfect squares as they should be.
Moreover, the cases (ii.) and (iii.) do not appear as
$(1_{NS}-1_R)$. This dissimilarity of the coefficients of the $NS$
and $R$ oscillators is exactly the effect of SS deformation and
the breaking of supersymmetry. All the amplitudes have the general
form
\bea 1_{NS}\left[ K_{NS,1} \sqrt{\cal V}_3+{K_{NS,2} \over
\sqrt{\cal V}_3}\right]^2 -1_{R}\left[ K_{R,1} \sqrt{\cal
V}_3+{K_{R,2} \over \sqrt{\cal V}_3}\right]^2~,
\label{K-general+SS}\eea
where $K_{NS,2} \sim (1+1)f(v_\a)$, $K_{R,2}\sim (1-1)f(v_\a)=0$
and $f(v_\a)$ a function of the shift vector $v_\a$. The second
term in both $K_{NS,2}$ and $K_{R,2}$ is due to $h$ twisted
states. This explains the appearance of the factor of 2 in the NS
sector in (\ref{khii}) and (\ref{khiii}) and the absence of the
factor proportional to $\frac{1}{\sqrt{\cal V}_3}$ in the R
sector.

\subsection*{Annulus}

To cancel these tadpoles one needs to add D9, D5$_3$ and
D5$_i$-branes as well as $\bar{\textrm{D}}5_i$-antibranes when
$R_i \in G$ with $i=1,2$, and the Scherk-Schwarz element $h$ acts
on the $T^2_3$ torus. The anti $\bar{\textrm{D}}5_i$-branes sit on
the $R_ih$ fixed points \cite{Scrucca:2001ni}.
\begin{center}
\epsfig{file=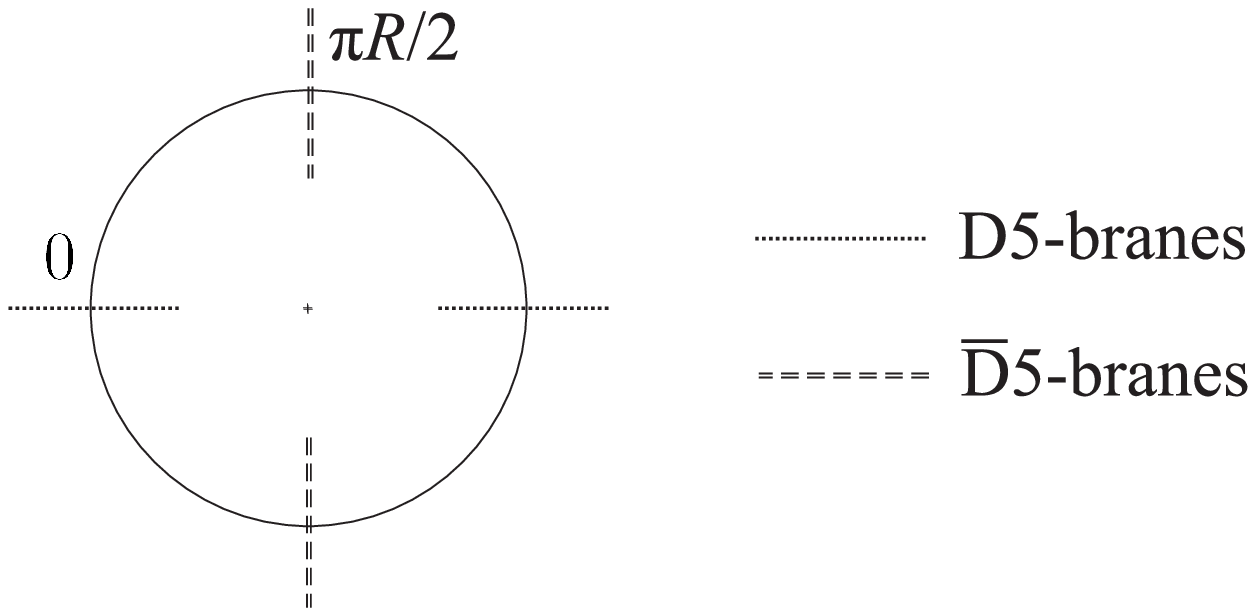,width=100mm}
\end{center}
The contribution from the Annulus amplitudes are the same as in
the supersymmetric case (without SS deformation) with in addition
the anti $\bar{\textrm{D}}5_{i\neq 3}$-brane sector when $R_i\in
G$. Note that the Annulus amplitudes between two D-branes
contributes $(1_{NS}-1_R)$, whereas, the ones between a D-brane
and an anti $\bar{\textrm{D}}$-brane leads $(1_{NS}+1_R)$
reflecting the breaking of supersymmetry. It is not difficult to
work out the contribution of an element $\a$ such that
$v_\a=(v_\a^1, v_\a^2, 0)$:
\begin{itemize}
\item[-] when $G$ contains no $Z_2$ factors\footnote{By $Z_2$
factors we refer to rotation elements and not to the
Scherk-Schwarz deformation $h$.} the contribution from the annulus
is the same as the one in the supersymmetric case (\ref{a}).
%
%
\item[-] when $G$ contains $Z_2$ factors, then:
\begin{itemize}
\item[i.] If only $R_3\in G$, the contribution is the same as the
one in the supersymmetric case (\ref{ai}).
%
%
\item[ii.] If only $R_{i}\in G$ for a given $i=1$ or $2$:
\bea
&& \frac{1}{\prod_{l} 2~\sin \pi v_\a^l}
\times\nonumber\\
&&\Bigg[~1_{NS}\bigg(\sqrt{{\cal V}_3}Tr[\gamma_{\a,9}]-
\frac{1}{\sqrt{{\cal V}_3}}2\sin\pi v_\a^j~ (Tr[\gamma_{\a,5_{i}}]
+Tr[\gamma_{\a,\bar{5}_{i}}])\bigg)^2
\nonumber\\
&&~-1_R\bigg(\sqrt{{\cal V}_3}Tr[\gamma_{\a,9}]-
\frac{1}{\sqrt{{\cal V}_3}}2\sin\pi v_\a^j~
\bigg(Tr[\gamma_{\a,5_{i}}]-Tr[\gamma_{\a,\bar{5}_{i}}]\bigg)
\bigg)^2\Bigg]~. \label{ahii} \eea
\item[iii.] If all possible $R_l\in G$, with $l=1,2,3$:
\bea &&\frac{1}{\prod_{l} 2\sin2\pi v_\a^l}
%
\Bigg[~1_{NS}\Bigg(\sqrt{{\cal V}_3} \bigg(Tr[\gamma_{\a,9}]+
\prod_{l} 2\sin\pi v_\a^l Tr[\gamma_{\a,5_3}]\bigg)
\nonumber\\
&& ~~~~~~~~~~~~~~~~~~~~- \frac{1}{\sqrt{{\cal V}_3}}\sum_{i\neq j
=1,2} 2\sin\pi v_\a^j \bigg(Tr[\gamma_{\a,5_i}]
+Tr[\gamma_{\a,\bar{5}_i}]
\bigg)\Bigg)^2 \nonumber\\
&&~~~~~~~~~~~~~~~~~ -1_R\Bigg(\sqrt{{\cal V}_3}
\bigg(Tr[\gamma_{\a,9}]+ \prod_{l} 2\sin\pi v_\a^l
Tr[\gamma_{\a,5_3}]\bigg)
\nonumber\\
&&~~~~~~~~~~~~~~~~~~~~-\frac{1}{\sqrt{{\cal V}_3}}\sum_{i\neq j
=1,2} 2\sin\pi v_\a^j \bigg(Tr[\gamma_{\a,5_i}]
-Tr[\gamma_{\a,\bar{5}_i}] \bigg)\Bigg)^2\Bigg]~. \label{ahiii}
\eea
\end{itemize}
\end{itemize}
%
%
The general form, is a function of the volume of the unaffected by
$\a$ torus (${\cal V}_3$):
\be 1_{NS}\left[ A_{NS,1} \sqrt{\cal V}_3+{A_{NS,2} \over
\sqrt{\cal V}_3}\right]^2 -1_{R}\left[ A_{R,1} \sqrt{\cal
V}_3+{A_{R,2} \over \sqrt{\cal V}_3}\right]^2~,
\label{A-general+SS}\ee
where $A_1$ and $A_2$ are again functions of the traces of the
Chan-Paton factors $Tr[\g_{\a, I}]$. $A_{NS,1}$ and $A_{R,1}$
are proportional to $Tr[\g_{\a,9}]$ and $Tr[\g_{\a,5_3}]$.
$A_{NS,2}$ and $A_{R,2}$ proportional to
$Tr[\g_{\a,5_i}]$ and $Tr[\g_{\a,\bar{5}_i}]$ for $i=1,2$.

\subsection*{M\"obius Strip}\label{Mobius+SS}

Finally, the M\"obius strip amplitude derived in
two inequivalent ways (as a direct amplitude and as
the mean value of the Klein Bottle and the Annulus amplitudes) leads the
same constraints as before (\ref{consta}) with in addition:
\begin{itemize}
\item[-] When $R_{i\neq 3}\notin G$, then:
\begin{itemize}
\item[i.] If $R_3\notin G$:
\bea Tr[\gamma^{T}_{\Omega h \a,9}\gamma^{-1}_{\Omega h \a,9}]=\pm
Tr[\gamma_{\a^2,9}]~, \eea
\item[ii.] If $R_3\in G$ we have in addition:
\bea Tr[\gamma^{T}_{\Omega h \a,5_3}\gamma^{-1}_{\Omega h
\a,5_3}]= \pm Tr[\gamma_{\a^2,5_3}] ~.\eea
\end{itemize}
We should take the same sign for the D9 and D5$_3$ sectors by
T-duality. Examples of this cases have been discussed in
\cite{Anastasopoulos:2003ha}.
\item[-] When $R_{i\neq 3}\in G$ for a given $i=1$ or $2$:
\bea Tr[\gamma^{T}_{\Omega h \a,I}\gamma^{-1}_{\Omega h
\a,I}]=Tr[\gamma_{\a^2,I}], ~~~~~~~~~I=9,5_3, 5_i,\bar{5}_i ~.\eea
In all cases $\gamma^2_{R,I}=-1$ with $I=9,5_l,\bar{5}_i$
for all $R$'s.
\end{itemize}
The details of the calculations are reported in the appendices.

\subsection*{Tadpole conditions:}\label{Tadpoles-WithSS}

The tadpole conditions for an element $\a^2$ such that
$v_\a=(v_\a^1,v_\a^2,0)$ are as follows:
\begin{itemize}
\item[-] When $G$ contains no $Z_2$ factors, the tadpole condition
is the same as in the case without SS deformation (\ref{tvk1}):
\bea Tr[\gamma_{\a^2,9}]= 32 \prod_l \cos \pi v_\a^l~.
\label{tvkh1} \eea
%
%
%
\item[-] When $G$ contains $Z_2$ factors:
\begin{itemize}
\item[i.] If only $R_3\in G$, the tadpole condition is the same as
in the case without SS deformation (\ref{tvk2i}):
\bea Tr[\gamma_{\a^2,9}]&+& 4\prod_{l}
\sin2\pi v_\a^l~Tr[\gamma_{\a^2,5_{3}}]
\nn\\
&=& 32~ \bigg(\prod_{l} \cos\pi v_\a^l + \prod_{l} \sin\pi
v_\a^l\bigg)~. \label{tvkh2i} \eea
\item[ii.] If only $R_{i}\in G$ for a given $i=1$ or $2$:
\bea
&& ~~~~~~~~~~~~~Tr[\gamma_{\a^2,9}]=32~\prod_{l} \cos\pi v_\a^l~, \nn\\
&& 1_{NS}: ~~~~2\sin2\p v_\a^j~\Big(Tr[\gamma_{\a^2,5_i}]+
Tr[\gamma_{\a^2,\bar{5}_i}]\Big)
= 32~\cos\pi v_\a^i~\sin\pi v_\a^j~,\nn\\
&& 1_R:~~~~~2\sin2\pi v_\a^j~\Big(Tr[\gamma_{\a^2,5_{i}}]-
Tr[\gamma_{\a^2,\bar{5}_{i}}]\Big)= 0~. \label{tvkh2ii} \eea
\item[iii.] If all possible $R_l\in G$ with $l=1,2,3$:
\bea
&&Tr[\gamma_{\a^2,9}]+4 \prod_{l}
\sin2\pi v_\a^l Tr[\gamma_{\a^2,5_3}]
=32~\Big(\prod_{l} \cos\pi v_\a^l + \prod_{l} \sin\pi
v_\a^l\Big)~,
\nn\\
&& 1_{NS}:~~~~~~~~2\sum_{i\neq j =1,2} \sin2\pi v_\a^j
\Big(Tr[\gamma_{\a^2,5_i}]+ Tr[\gamma_{\a^2,\bar{5}_{i}}]\Big) \nn\\
&&~~~~~~~~~~~~~~~~~~~~~~~~~~~~~~~~~~~~ =32~\sum_{i\neq j =1,2}
\epsilon_{ij}~\cos\pi v_\a^i \sin\pi v_\a^j~,
\nn\\
&& 1_{R}:~~~~~~~~~2\sum_{i\neq j =1,2} \sin2\pi v_\a^j
\Big(Tr[\gamma_{\a^2,5_i}]-Tr[\gamma_{\a^2,\bar{5}_{i}}]\Big)=0~.
\label{tvkh2iii} \eea
\end{itemize}
\end{itemize}

Finally, there could be elements that can not be expressed as the
square of any other element in $G$. These elements will not
receive contribution from the Klein Bottle amplitude. For such
elements and for elements of the form $R_3\a$ the tadpole
conditions are the same as (\ref{tvkh1}-\ref{tvkh2iii}) with zero
on the right hand side. For the group elements $h\a$ and $R_3h\a$
the tadpole conditions are as (\ref{tvkh1}) and (\ref{tvkh2i}) and
since D5$_{i}$ and $\bar{\textrm{D}}5_{i}$ are transverse to the
direction where $h$ acts, there are no conditions on
$Tr[\gamma_{h\a,5_{i}}]$ and $Tr[\gamma_{h\a,\bar{5}_{i}}]$. For
the group element $R_i\a$ the tadpole condition is
\bea Tr[\gamma_{R_i\a,9}]&+& 4\sin\pi v_\a^i \cos\pi v_\a^j
Tr[\gamma_{R_i\a,5_3}]
\nonumber\\
&+&2\cos\pi v_\a^j Tr[\gamma_{R_i\a,5_i}]+2\sin\pi v_\a^i
Tr[\gamma_{R_i\a,5_j}]=0~, \eea
where the $\bar{\textrm{D}}5_i$-branes do not contribute because
they do not sit on the fixed points of this element. This
condition is valid for both NS and R sectors (multiplied by
$(1_{NS}-1_R)$). For the element $R_ih\a$ we find
\bea 1_{NS}:~~Tr[\gamma_{R_ih\a,9}]&+& 4\sin\pi v_\a^i \cos\pi
v_\a^j Tr[\gamma_{R_ih\a,5_3}]
\nn\\
&+&2\cos\pi v_\a^j Tr[\gamma_{R_i\a,\bar{5}_i}]
+2\sin\pi v_\a^i Tr[\gamma_{R_ih\a,\bar{5}_j}]=0~,\nn\\
1_{R}:~~Tr[\gamma_{R_ih\a,9}]&+& 4\sin\pi v_\a^i \cos\pi v_\a^j
Tr[\gamma_{R_ih\a,5_3}]
\nn\\
&-&2\cos\pi v_\a^j Tr[\gamma_{R_ih\a,\bar{5}_i}] -2\sin\pi v_\a^i
Tr[\gamma_{R_ih\a,\bar{5}_j}]=0~, \eea
where the D5$_i$-branes do not contribute because they do not sit
on the fixed points of $R_ih\a$.



\section{Some specific examples}\label{Examples}

Let us discuss some examples that can be described by the general
formulae we have obtained in the previous sections.

The first example is the orbifold groups studied by Gimon and
Johnson \cite{Gimon:1996ay}, where $G=Z_N$ for $N=2,3,4,6$ acting
on $T^4$. The tadpole conditions are given by
(\ref{tvkh1}-\ref{tvkh2i}) with
$v_\a^1=-v_\a^2=\frac{k}{N},~v_\a^3=0$ leading for odd $N$
\bea Tr[\gamma_{2k,9}]= 32 \cos^2\frac{k}{N}\pi~, \eea
whereas for even $N$:
\bea
&&Tr[\gamma_{2k,9}]-4~\sin^2\frac{2k}{N}\pi~Tr[\gamma_{2k,5_{3}}]=
32~ \cos\frac{2k}{N}\p~,
\nn\\
&&Tr[\gamma_{2k-1,9}]-4~\sin^2\frac{2k-1}{N}\pi~Tr[\gamma_{2k-1,5_{3}}]=
0~. \eea
The tadpole conditions for the groups studied by Zwart
\cite{Zwart:1997aj}, where $G=Z_N\times Z_M$, can be easily
reproduced by using (\ref{tvk1}-\ref{tgiv}). As an example the
tadpole conditions for $Z_N\times Z_M$ where both $N$ and $M$ odd
are given by:
\bea
Tr[\g_{2k,0,9}]&=&32~\cos^2\frac{k}{N}\p~, \nn\\
Tr[\g_{0,2l,9}]&=&32~\cos^2\frac{l}{M}\p~, \nn\\
Tr[\g_{2k,2l,9}]&=& 32~\cos\frac{k}{N}\p~\cos\frac{l}{M}\p~
\cos\bigg(\frac{l}{M}-\frac{k}{N}\bigg)\p~, \eea
all the other cases can be easily worked out in a similar way.
The tadpole conditions for the groups studied by Ibanez et al
\cite{Aldazabal:1998mr}, where $G=Z_N$ with both $N$ even and odd
and $G=Z_N\times Z_M$ with $N$ or $M$ even, can be reproduced
using (\ref{tvk1}-\ref{tgiv}) by specifying the vector $v_\a$ for
the different group elements.

In \cite{Zwart:1997aj, Aldazabal:1998mr} it was found that the
tadpole conditions for $Z_4$, $Z_8$, $Z'_8$ and $Z'_{12}$ have no
solutions. In particular, for $Z_4$ with
$v_\a=(\frac{1}{4},\frac{1}{4},-\frac{1}{2})$ the tadpole
conditions can be written as:
\bea Tr[\g_{\a,9}]+ 2 Tr[\g_{\a,5_3}]&=& 0~,
\nn\\
Tr[\g_{\a^2,9}]+ 4 Tr[\g_{\a^2,5_3}]&=& 0~,
\nn\\
Tr[\g_{\a^3,9}]+ 2 Tr[\g_{\a^3,5_3}]&=& 0~, \label{Z4} \eea
plus a contribution in the Klein Bottle of the form (\ref{kgv1})
proportional to $1/{\cal V}_3$ that can not be cancelled by any
D-brane contribution.
This is a general feature for all the inconsistent groups. They
contain elements $\a$ of the form $\a=\b R_i$ where
$v_\b=(v_\b^1,v_\b^2,0)$ and $R_i\notin G$ contributing a
$\frac{1}{{\cal V}_3}$ to the tadpoles. Therefore, the
introduction of D9 and D5$_3$-branes contributing ${\cal V}_3$
alone is not enough to cancel all the tadpoles. In
\cite{Kakushadze:1998cd} it has been argued that these
orientifolds contain non-perturbative sectors that are missing in
the world-sheet approach. Attempts to cancel these tadpoles has
been given in several papers \cite{sagn, Pradisi:1988xd,
Angelantonj:2002ct, Rabadan:2000ma, Klein:2000tf} by adding
discrete torsion or Wilson lines.
In the next section we will solve the tadpole conditions for the
non-supersymmetric version of $Z_4$, where supersymmetry is broken
by an order two Scherk-Schwarz deformation and show that the
obtained spectrum is chiral and anomaly free.

\section{Solving the tadpole conditions}\label{Solving}

Open string states are denoted by $|{\psi},ab\rangle$, where
${\psi}$ refers to the world-sheet degrees of freedom while $a$,
$b$ are Chan-Paton indices associated to open string endpoints
lying on D-branes. The action of a group element $\a$ on this
state is given by
\bea \a: |{\psi},ab\rangle ~~~\to ~~~(\g_\a)_{a,a'} |\a
{\psi},a'b'\rangle (\g_\a)^{-1}_{b'b}~,
\eea
where $\g_\a$ is a unitary matrix associated to $\a$. the action
of $\Omega \a$ is given by
\bea \Omega \a: |{\psi},ab\rangle ~~~\to ~~~(\g_{\Omega
\a})_{a,a'} |\a {\psi},b'a'\rangle (\g_{\Omega \a})^{-1}_{b'b}~,
\eea
The Chan-Paton matrices must be such that the full states are invariant
under the action of the orientifold group. Hence,
\bea
\l^{(i)}~\sim~ e^{2\pi i b_\a^i} ~~\g_a \l^{(i)} \g_\a^{-1}
~,~~~~~~~\l^{(i)} ~\sim ~\g_\Omega \l^{(i)}{}^T \g_\Omega^{-1}~.
\label{lambda}
\eea
A simple way to impose the tadpole conditions on the Chan-Paton
matrices $\l$ is to recast them in a Cartan-Weyl basis. In this
case, constraints on $\l$'s will emerge as restrictions on weight
vectors \cite{Aldazabal:1998mr}.
They can be organized into charged generators: $\l_a=E_a$ and Cartan
generators $\l_I=H_I$ such that:
\bea [H_I,E_a]=\r_I^a E_a~,\label{Cartan}\eea
where $\r_I^a=(\underline{\pm1,\pm1,0,\cdots,0})$ are the roots
associated to the generators $E_a$ of the $SO(32)$ Lie-algebra.
The underline denotes all the possible permutations. The matrix
$\g_{\a}$ and its powers represent the action of the orientifold
group on the Chan-Paton factors, and they correspond to elements
of a discrete subgroup of the Abelian group spanned by the Cartan
generators. Hence, we can write:
\bea \g_{\a}=e^{-2\p i V_\a\cdot H}~,\label{ShiftVector}\eea
where the dimension of the $shift$ vector $V_\a$ is the rank of
the $SO(32)$ Lie-group. Cartan generators are represented by
$2\times 2$ $\sigma_3$ submatrices.
%

Recalling the formula
$e^{-B} A e^B=\sum_{n=0}^\infty [A,B]_n$
with $[A,B]_{n+1}=[[A,B]_{n},B]$ and $[A,B]_0=A$, and using
(\ref{Cartan}), it is easy to show that:
\bea \g_\a E_a \g_\a^{-1}=e^{-2\p i \r_a\cdot V_\a} E_a~,\eea
The equations giving the massless spectrum (\ref{lambda}) can be
expressed in the following way:
\bea \r_I^a \cdot V_\a= b_\a~,
%
\label{Vtad}
\eea
where ``$b_\a$" is associated with the transformation of the
corresponding massless fields. Notice the difference between 99,
5$_i$5$_i$ and 95$_i$ strings:
%
%
%
\bea b_{99,5_i5_i}= \left\{ \ba{lcl} 0 && \textrm{for vectors}\\
                            v_\a^i && \textrm{for scalars}\\
                            s\cdot v_\a && \textrm{for fermions}
                            \ea \right.
%
%
~,~
%
b_{95_i}= \left\{ \ba{lcl} s_jv_j+s_lv_l && \textrm{for scalars}\\
                            s_i v_i && \textrm{for fermions}
                            \ea \right. \label{Shifts}\eea
%
where $b_\a$'s should be understood modulo $\Zint$.

In general, the matrix $\g_{\a,p}$ which acts on a given Dp-brane,
satisfy:
\be \g_\a^k=\pm 1~, \label{order} \ee
where $k$ is the order of the associated orientifold group element
$\a$, and it is the smallest integer such that $\a^k=1$. Depending
on the sign choice in (\ref{order}), the vector $V_\a$ has the
form
\bea
V_\a={1\over k}\Big( 0^{n_0},1^{n_1},\cdots,j^{n_j}, \cdots
\Big) ~,\label{Vodd}\eea
for the minus sign, where $j=0, \cdots, k-1$ and
\bea V_\a={1\over
2k}\Big(1^{n_1}, 3^{n_2},\cdots, (2j-1)^{n_j},
\cdots
\Big)~, \label{Veven}\eea
for the plus sign, where $j=0,\cdots, {k\over 2}$. The number of
each entry is determined by the corresponding tadpole condition.

The $(pq)$ sector is handled using an auxiliary $SO(64)\supset
SO(32)_{(pp)}\otimes SO(32)_{(qq)}$ algebra. Since we have
generators acting simultaneously on both Dp and Dq branes, only
roots of the form:
\bea \r_{(pq)}=\r_{(p)}\otimes \r_{(q)}=(\underline{\pm 1,
0,\ldots,0};\underline{\pm 1, 0,\ldots,0})~,\eea
must be considered. The shift vector is than defined as
\be W_{(pq)}=V_{(p)}\otimes V_{(q)}~. \ee
In \cite{Aldazabal:1999nu} the full anomaly free open string
spectrum for odd and even order (with one $Z_2$ factor)
orientifolds was found for a twist vector given by
$\frac{1}{N}(t_1,t_2,t_3)$ where $\sum_i t_i=0$. We would like to
study orientifold models with more than one $Z_2$ namely two $Z_2$
factors and show that we can use the same techniques for most
cases, namely $Z_2\times Z_2\times G$ where $G$ is a group with no
$Z_2$ elements and work out the open string spectrum.

\subsection*{Supersymmetric $Z_2\times Z_2\times Z_N$}


The supersymmetric $Z_2\times Z_2$ orientifold model was solved in
\cite{Berkooz:1996dw} where the gauge group was found to be
$USp(16)$ with scalars in the antisymmetric representation
$\Yasymm = 120$.

Following \cite{Aldazabal:1999nu} we will be more general and
consider also $Z_N$ orbifold actions that are not necessarily
crystallographic. A closed inspection of the tadpole conditions
shows that for $N$ odd the $Z_N$ matrices $\g_\a$ commutes with
all the $Z_2$ matrices $\g_{R_i}$, therefore, to solve the tadpole
conditions we can use the Cartan-Weyl basis. For the spectrum we
have:
%
%
\begin{itemize}
\item Gauge bosons:
\be \r_I^a \cdot V_\a= 0~, \ee
where now $\r_I^a$ are the eight dimensional roots associated with
the generators of $USp(16)$ Lie algebra. These are
$\r_I^a=(\underline{\pm 1, \pm 1,0,\cdots,0})$ together with the
long roots $\r_I^a=(\underline{\pm 2, 0, \cdots,0})$.
\item Scalars:
\be \r_I^a \cdot V_\a= b_\a ~,\ee
where $b_\a$ is given in (\ref{Shifts}) and $\r_I^a$ are the roots
associated with the generators of the antisymmetric representation
of $USp(16)$, namely, $\r_I^a=(\underline{\pm 1, \pm 1,0,
\cdots,0})$.
\end{itemize}


\subsubsection*{General Massless formulae}

Using the technic described above, we can work out the general
massless formulae for the case $Z_2\times Z_2 \times Z_N$ with $N$
odd and shift vector acting as $(v_\a^1,v_\a^2,0)$ \footnote{The
$n_i$'s are defined by the tadpole conditions.}:
\begin{itemize}
\item For 99/55 states, we have:
\begin{itemize}
\item[$\triangleright$] Vectors in: $USp(n_1)\times
\prod^{(N+1)/2}_{i=2}U(n_i)$.
\item[$\triangleright$] Scalars $\y_{-1/2}^I|0\rangle$ in:
$(n_i,n_{2-i+b^I_{\a} N})$, $(n_i,\overline{n}_{i-b^I_{\a} N})$,
$(\overline{n}_i,\overline{n}_{2-i-b^I_{\a} N})$.

Representations of $USp(n_1)$ appear as $n_1+\bar{n}_1$ that
denote the vector $n_{1,v}$.
There will be antisymmetric reps in the $U(n_i)$ iff $2i-2=v_I N$.
\end{itemize}
%
%
%
%
\item For the $95_i$/$5_j5_k$ states we have:
\begin{itemize}
\item[$\triangleright$] Scalars $|s_1,s_2\rangle$
(95$_3$/$5_15_2$) in: $(n_1,\tilde{n}_1)$,
$(n_i,\overline{\tilde{n}}_i)$, $(\overline{n}_i,\tilde{n}_i)$,

\item[$\triangleright$] Scalars $|s_1,s_3\rangle$
(95$_2$/$5_15_3$) in:
$(n_i,\tilde{n}_{2-i+b^2_{\a} N})$, $(n_{2-i+b^2_{\a} N},
\tilde{n}_i)$,
$(n_i,\overline{\tilde{n}}_{i-b^2_{\a} N})$,
$(\overline{n}_{i-b^2_{\a} N}, \tilde{n}_i)$,
$(\overline{n}_i,\overline{\tilde{n}}_{2-i-b^2_{\a} N})$,
$(\overline{n}_{2-i-b^2_{\a} N}, \overline{\tilde{n}}_i)$.

\item[$\triangleright$] Scalars $|s_2,s_3\rangle$
(95$_1$/$5_25_3$) in:
$(n_i,\tilde{n}_{2-i+b^1_{\a} N})$, $(n_{2-i+b^1_{\a} N},
\tilde{n}_i)$,
$(n_i,\overline{\tilde{n}}_{i-b^1_{\a} N})$,
$(\overline{n}_{i-b^1_{\a} N}, \tilde{n}_i)$,
$(\overline{n}_i,\overline{\tilde{n}}_{2-i-b^1_{\a} N})$,
$(\overline{n}_{2-i-b^1_{\a} N}, \overline{\tilde{n}}_i)$.
\end{itemize}
\end{itemize}
Fermions will transform in similar representations due to
supersymmetry (up to now we discuss only the action of rotation
elements onto the Chan-Paton factors).

\subsection*{Scherk-Schwarz deformation}

Scherk-Schwarz deformation acts on the open strings in the same
way as the rotation elements do. The shift vector corresponding to
$\g_h^2=\pm 1$ is given by:
\bea V_h={1\over 4} \left\{ \ba {lcl}
(1_a,-1_b) & \quad & \textrm{for $\g^2_h=-1$}\\
%
(2_a,0_b)& \quad & \textrm{for $\g^2_h=+1$} \ea \right.
\label{V-h}\eea
where the index refer to the number of components in
the vector.
%
The action of the Scherk-Schwarz deformation is as (\ref{lambda}) with:
\bea b_{h}= \left\{\ba{lcl} 0 && \textrm{for spacetime bosons}\\
                            1/2 && \textrm{for spacetime fermions}
                            \ea \right.
\label{Shifts-h}\eea
where $b_h$ is defined modulo $\Zint$.

When the SS deformation acts on a supersymmetric model, it breaks
the gauge group for $\g^{2}_h=-1$, as:
\be U(N)\rightarrow U(n)\times U(N-n)~,\ee
%
and for both $G=SO(2N), USp(2N)$
\be G\rightarrow U(N)~,\ee
whereas for $\g^{2}_h=+1$ as:
\bea G_N\rightarrow G_n\times G_{N-n}~,
\eea
where $G_n=U(n), SO(n)$ and $USp(n)$. The remaining
representations split accordingly for $\g^{2}_h=-1$:
\bea
(m,n) \to\left\{\ba{lcl} (m_1,n_2)+(m_2,n_1) && \textrm{bosons}\\
                            (m_1,n_1)+(m_2,n_2) && \textrm{fermions}
                            \ea \right.\eea
%
%
%
whereas for $\g^{2}_h=+1$:
\bea(m,n) \to\left\{\ba{lcl} (m_1,n_1)+(m_2,n_2) && \textrm{bosons}\\
                            (m_1,n_2)+(m_2,n_1) && \textrm{fermions}
                            \ea \right.\eea
and for both $\g^{2}_h=\pm 1$ the bifundamental representations
split as:
\bea(m,\bar{n}) \to\left\{\ba{lcl}
(m_1,\bar{n}_1)+(m_2,\bar{n}_2)
                            && \textrm{bosons}\\
                            (m_1,\bar{n}_2)+(m_2,\bar{n}_1)
                            && \textrm{fermions}\ea \right.\eea
where $m=m_1+m_2$ and $n=n_1+n_2$.

\subsection*{Non-Supersymmetric $Z_4$}

Consider the non-supersymmetric version of $Z_4$ orientifold by
adding a SS deformation. The orbifold group is generated by a
twist whose action on the three complex coordinates is given by
$v_\a=(\frac{1}{4},\frac{1}{4},-\frac{1}{2})$. The undesired
aforementioned Klein-bottle R-R tadpoles cancel between
(\ref{kgv1}) and (\ref{kghv1}). This is true also for $Z_8$,
$Z'_8$ and $Z_{12}$. The remaining tadpole conditions are the same
as before (\ref{Z4}) with the same conditions for $h\a$ where
$\g^2_{h,9}=\g^2_{h,5_3}=+1$. Their cancellation requires
$Tr\g_\a=Tr\g_{h\a}=0$ for all elements $\a$ of $Z_4$. Hence,
%
%
%
%
They can be easily solved with the matrices:
%
%
%
\bea
\g_\a&=&\textrm{diag}[\f {\bf I}_a, \f^{-1} {\bf I}_a, \f {\bf
I}_b, \f^{-1} {\bf I}_b, \f^3 {\bf I}_c, \f^{-3} {\bf I}_c, \f^{3}
{\bf I}_d, \f^{-3} {\bf I}_d]~,
\nn\\
\g_h&=&\textrm{diag}[-{\bf I}_a, -{\bf I}_a, {\bf I}_b, {\bf I}_b,
-{\bf I}_c, -{\bf I}_c, {\bf I}_d, {\bf I}_d]~,
\eea
where $a+b=c+d=8$, $\f=e^{2 \pi i/8}$ and $\g_\a^4=-{\bf I}$. The
open string massless spectrum is given in Table \ref{Z-4}, it is
chiral and does not suffer of irreducible gauge anomalies.

Another non-supersymmetric version of this model can be obtained
by using the projection $\Omega^{\prime}=\Omega \cdot J$ (instead
of $\Omega$), which gives a relative sign between the untwisted
and twisted states in the Klein bottle \cite{Angelantonj:2002ct}.
This action is called ``discrete torsion" and it introduces $O^+$
orientifold planes instead of $O^-$. To cancel the resulting R-R
tadpoles the introduction of D9 and anti
$\bar{\textrm{D}}5$-branes (instead of D5-branes) are needed. In
terms of Chan-Paton matrices this action is reflected as a
constraint on the $\g$'s by requiring them to satisfy
$\g_\a^4=+{\bf I}$.
%
%
The matrices which satisfy all the tadpole conditions are:
\bea
\g_\a&=&\textrm{diag}[i {\bf I}_a, -i {\bf I}_a, i {\bf I}_b, -i
{\bf I}_b, -{\bf I}_c, -{\bf I}_d, {\bf I}_e, {\bf I}_f]
\noN\\
\g_h&=&\textrm{diag}[-{\bf I}_a, -{\bf I}_a, {\bf I}_b, {\bf I}_b,
-{\bf I}_c, {\bf I}_d, -{\bf I}_e, {\bf I}_f]
\eea
where $c+d=e+f$. The spectrum is again non-chiral and anomaly free
(Table \ref{Z-4}).
\begin{table}[h]\footnotesize \renewcommand{\arraystretch}{.8}
\begin{tabular}{|c c c|}
\hline \hline
& & \\
& \raisebox{.8ex}[0cm][0cm]{\textbf{Z}$_4\times$
\textbf{SS}$~~~~(\g_\a^4=-{\bf I})$} &  \\
\hline \hline \hline
$U(a)\times U(b)\times$& & \\
$U(c)\times U(d)_{9,5}$ &
\raisebox{.8ex}[0cm][0cm]{(99)/(55)~matter}
& \raisebox{.8ex}[0cm][0cm]{(95)~matter}\\
\hline \hline
& & \\
& $2 \left( (\bar{a},c)+(\bar{b},d)+ \YasymmS_a + \YasymmS_b+
\bYasymmS_c+ \bYasymmS_d \right) + $
& $(a;a)+ (\bar{a};c)+ (b;b)+ (\bar{b};d)+ $ \\
\raisebox{.8ex}[0cm][0cm]{Bosons} &
$(\bar{a},\bar{c})+ (a,c)+(\bar{b},\bar{d})+(b,d)$ &
$(\bar{c};\bar{c})+ (c;\bar{a})+ (d;\bar{b})+ (\bar{d};\bar{d})$\\
& & \\
\hline
& & \\
& $(a,\bar{b})+ (\bar{c},d)+ c.c. +$&
$(a;b)+ (\bar{a};d)+ (b;a)+ (\bar{b};c)+$\\
Fermions & $2 \left( (a,b)+ (\bar{a},d)+ (\bar{b},c)+
(\bar{c},\bar{d}) \right)$
& \\
& $(a,d)+ (\bar{a},\bar{d})+ (b,c)+ (\bar{b},\bar{c})$ &
$(c;\bar{b})+ (\bar{c};\bar{d})+ (d;\bar{a})+ (\bar{d};\bar{c})$\\
& & \\
\hline
%
%
%
%
%
%
\hline \hline
& & \\
& \raisebox{.8ex}[0cm][0cm]{\textbf{Z}$_4\times$
\textbf{SS}$~~~~(\g_\a^4=+{\bf I})$} &  \\
\hline \hline \hline
$U(a)\times U(b)\times SO(c)$& & \\
$SO(d)\times SO(e)\times SO(f)_{9,5}$ &
\raisebox{.8ex}[0cm][0cm]{(99)/(55)~matter}
& \raisebox{.8ex}[0cm][0cm]{(95)~matter}\\
\hline \hline
& & \\
& $2 \left((\bar{a},c)+(a,e)+
(\bar{b},d)+(b,f) \right) + $
& $(\bar{a};d)+ (a;f)+ (\bar{b};c)+ (b;e)+$ \\
\raisebox{.8ex}[0cm][0cm]{Bosons} &
$\YasymmS_a+ \bar{\YasymmS}_a+ \YasymmS_b+ \bar{\YasymmS}_b+
(c,e)+ (d,f)$ &
$(c;\bar{b})+ (d;\bar{a})+ (e;b)+ (f;a)$\\
& & \\
\hline
& & \\
& $(a,\bar{b})+ (\bar{a},b)+ (c,d)+ (e,f)+$&
$(\bar{a};d)+ (a;f)+ (\bar{b};c)+ (b;e)+$\\
Fermions & $2 \left( (\bar{a},d)+ (a,f)+ (\bar{b},c)+ (b,e)
\right)$
& \\
& $(a,b)+ (\bar{a},\bar{b})+ (c,f)+ (d,e)$ &
$(c;\bar{b})+ (d;\bar{a})+ (e;b)+ (f;a)$\\
& & \\
\hline
\end{tabular}
\caption{The massless spectrum of the non supersymmetric $Z_4$
models.}\label{Z-4}
\end{table}
%
%

\subsection*{Non-Supersymmetric $Z_6\times Z_2$}

Another example that we consider is the non-supersymmetric version
of the $Z_6\times Z_2$ orientifold by SS deformation. The
supersymmetric version was discussed by Zwart \cite{Zwart:1997aj}.
%
%
%
%
The shift vectors of the different elements are:
\bea
v_\a={1\over 3} (-1,1,0) ~,~~~ g_1={1\over 2} (0,-1,1) 
~,~~~ g_2={1\over 2} (1,0,-1)~.
\eea
where there is another $Z_2$ element with shift vector
$g_3=g_1+g_2$.
The $\Omega$ action on the D9-branes is described by a symmetric
matrix that we choose to be:
\be \g_{\Omega,9}={\bf I}_{32}~. \ee
Supersymmetry is broken by introducing a SS deformation $h$ acting
on $T^2_3$ with a shift of order two. By studying the Klein bottle
amplitudes we realize that there are two $O^-$-planes sitting on
the $R_3=R_1\cdot R_2$ fixed points. The $R_1$ and $R_2$ elements
act in the same direction as the SS deformation $h$ giving
$O^-$-planes sitting on the two $R$-fixed points and
$\bar{O}^-$-planes on the $R h$-fixed points. To cancel the
tadpoles we need to add D5$_i$-branes on the $R_i$-fixed points
with $i=1,2,3$ and $\bar{\textrm{D}}5_i$ antibranes on the $R_i
h$-fixed points with $i=1,2$.
The $\g_{R_l}$ matrices satisfying the tadpole conditions, are:
\bea
\g_{R_1,9}=
\left( \ba {cccc}
 0       &-{\bf I} \\
 {\bf I} & 0  \ea \right)~,~~~
\g_{R_2,9}=i
\left( \ba {cccc}
 \e & 0 \\
 0 & -\e \ea \right)~,~~~
\g_{R_3,9}=-i
\left( \ba {cccc}
 0 & \e \\
 \e & 0 \ea \right)
\eea
where $\e={\bf I}_{2} \otimes \s_2 \otimes {\bf I}_{4}$. The $Z_3$
action on the Chan Paton matrices is given by the reducible
matrix:
\be \g_{\a,9}= \textrm{diag}[A_4, A_4, {\bf I}_8, A_4, A_4, {\bf
I}_8] \ee
where $A_4={\bf I}_2\otimes \left( \ba {cccc}
-1/2 & -\sqrt{3}/2 \\
\sqrt{3}/2 & -1/2 \ea \right)$
and SS action on the Chan-Paton is given by:
\be \g_{h,9}= \textrm{diag}[{\bf I}_4, -{\bf I}_8, {\bf I}_8,
-{\bf I}_8, {\bf I}_4] \ee
where $\g_{h,9}^2=+{\bf I}$.
The gauge group for the D9 branes is $U(2)\times U(2)\times
USp(4)\times USp(4)$.

For the D5 branes, $\g_\Omega$ is antisymmetric, therefore, we can
choose:
\be \g_{\Omega,5_i}=i\s_2 \otimes {\bf I}_{16}~. \ee
for all $i=1,2,3$. The orbifold action is then:
\bea
\g_{R_i,5_i}=(-1)^{\d_{3,i}}
\left( \ba {cccc}
 0 &-{\bf I} \\
 {\bf I} & 0  \ea \right)~,~~~
\g_{R_i,5_{i+1}}=
\left( \ba {cccc}
 {\bf I} & 0 \\
 0 &-{\bf I} \ea \right)~,~~~
\g_{R_i,5_{i+2}}=
\left( \ba {cccc}
 0 &-{\bf I} \\
-{\bf I} & 0 \ea \right)~
\eea
where the indexes are defined modulo 3. The $Z_3$ action on the
Chan Paton is given by:
\be \g_{\q,5_i}= (-1)^{\d_{2,i}}\textrm{diag}[\f {\bf I}_2, \f^{-1} {\bf I}_2,
\f{\bf I} _2, \f^{-1} {\bf I}_2, {\bf I}_8, \f {\bf I}_2,
\f^{-1} {\bf I}_2, \f {\bf I}_2, \f^{-1} {\bf I}_2, {\bf I}_8]~. \ee
where $\f=e^{2\p i/3}$. The SS action on the D5$_i$ Chan Paton
matrices are the same as the D9 one $\g_{h,5_i}=\g_{h,9}$.
\begin{table}
\footnotesize \renewcommand{\arraystretch}{1.25}
\begin{tabular}{|c c c|}
\hline \hline
& & \\
& \raisebox{2.5ex}[0cm][0cm]{\textbf{Z}$_6\times$\textbf{Z}$_2$+SS} &  \\
\hline \hline \hline
& & \\
\raisebox{2.5ex}[0cm][0cm]{ Sectors }&
\raisebox{2.5ex}[0cm][0cm]{ Bosons }&
\raisebox{2.5ex}[0cm][0cm]{ Fermions }
\\
\hline \hline
$99$/ $5_35_3$ & $U(2)\times U(2)\times USp(4)\times USp(4)$
&2 ($(\bar{2},2,1,1)$+ $(2,\bar{2},1,1)$+ $(1,1,4,4)$)\\
&$\bYasymmS_a$+ $\bYasymmS_b$+ $(2,1,1,4)$+ $(1,2,4,1)$
&$(2,2,1,1)$+ $(\bar{2},1,4,1)$+ $(1,\bar{2},1,4)$\\
&$\YasymmS_a$+ $\YasymmS_b$+ $(\bar{2},1,1,4)$+ $(1,\bar{2},4,1)$&
$(\bar{2},\bar{2},1,1)$+ $(2,1,4,1)$+ $(1,2,1,4)$\\
&+Scalars in the adjoints&\\
%
%
\hline \hline
$5_15_1$/ $\bar{5}_1 \bar{5}_1$/
&$U(4)\times USp(8)$&\\
$5_25_2$/ $\bar{5}_2 \bar{5}_2$&$(\YasymmS,1)$+
$(\bYasymmS,1)$+$(\bar{4},8)$ + $(4,8)$&
$(\YasymmS,1)$+$(\bYasymmS,1)$ + $(\bar{4},8)$+ $(4,8)$\\
(SUSY)
&Scalars in the Adjoint& 2 Fermions in the Adjoint\\
%
%
\hline \hline
$95_1$/ $5_35_2$& $(\bar{2},1,1,1;\bar{4},1)$+ $(2,1,1,1;1,8)$+
&$(1,2,1,1;4,1)$+ $(1,\bar{2},1,1;1,8)$+
\\
&$(1,1,1,4;4,1)$&$(1,1,4,1;\bar{4},1)$\\
\hline \hline
$9\bar{5}_1$/ $5_3\bar{5}_2$ &$(\bar{2},1,1,1;\bar{4},1)$+
$(2,1,1,1;1,8)$+
&$(1,2,1,1;4,1)$+ $(1,\bar{2},1,1;1,8)$+\\
&$(1,1,1,4;4,1)$&$(1,1,4,1;\bar{4},1)$\\
%
%
\hline \hline
$95_2$/ $5_35_1$ &$(2,1,1,1;4,1)$+ $(\bar{2},1,1,1;1,8)$
&$(1,\bar{2},1,1;\bar{4},1)$+ $(1,2,1,1;1,8)$\\
&+$(1,1,1,4;\bar{4},1)$&+$(1,1,4,1;4,1)$\\
\hline \hline
$9\bar{5}_2$/ $5_3\bar{5}_1$ &$(2,1,1,1;4,1)$+
$(\bar{2},1,1,1;1,8)$
&$(1,\bar{2},1,1;\bar{4},1)$+ $(1,2,1,1;1,8)$\\
&+$(1,1,1,4;\bar{4},1)$&+$(1,1,4,1;4,1)$\\
\hline \hline
$95_3$&$(2,1,1,1;\bar{2},1,1,1)$+
$(\bar{2},1,1,1;2,1,1,1)$
&$(2,1,1,1;1,\bar{2},1,1)$+
$(\bar{2},1,1,1;1,2,1,1)$\\
&$(1,2,1,1;1,\bar{2},1,1,1)$+ $(1,\bar{2},1,1;1,2,1,1)$
&$(1,2,1,1;\bar{2},1,1,1)$+
$(1,\bar{2},1,1;2,1,1,1)$\\
&$(1,1,4,1;1,1,4,1)$+ $(1,1,1,4;1,1,1,4)$&$(1,1,4,1;1,1,1,4)$+
$(1,1,1,4;1,1,4,1)$\\
\hline \hline
$5_15_2$/ $\bar{5}_1\bar{5}_2$
&$(4,1;\bar{4},1)$+ $(\bar{4},1;4,1)$
&$(4,1;\bar{4},1)$+ $(\bar{4},1;4,1)$\\
(SUSY) & $ (1,8;1,8)$&$(1,8;1,8)$\\
\hline \hline
\end{tabular} \caption{The massless spectrum of the
$Z_6\times Z_2$ with a Scherk-Schwarz deformation. The spectrum is
chiral and anomaly free. Notice the supersymmetric and not
supersymmetric sectors.}\label{Z-6}
\end{table}

The spectrum is chiral and anomaly free (Table \ref{Z-6}). Few
comments are in order:
a) Since $h$ acts longitudinal to D9 and D5$_3$ branes, their
corresponding open strings are affected by the action of the SS
deformation. Hence, the spectrum appears to be non-supersymmetric.
b) Moreover, since $h$ acts transverse to D5$_1$, D5$_2$,
$\bar{\textrm{D}}5_1$ and $\bar{\textrm{D}}5_2$ their
corresponding open strings are not affected by the SS deformation.
The spectrum in these cases is supersymmetric.
c) Notice the T-dualities exchanging $\textrm{D}9 \leftrightarrow
\textrm{D}5_3$, $\textrm{D}5_1\leftrightarrow \textrm{D}5_2$ and
$\bar{\textrm{D}}5_1\leftrightarrow \bar{\textrm{D}}5_2$ branes.
The open string spectrum that we found does indeed obey these
duality symmetries.
%
%

%
%
%
%
%
%
%
%
%
%
%
%
%
%
%
%
%
%
%
%
%
%
%
%
%
%

In this chapter we analytically solve the tadpole conditions and
we evaluate the massless spectrum of the $Z_4$ and the $Z_6\times
Z_2$ orientifolds with a Scherk-Schwarz deformation. However,
another way to obtain the same solutions is by using the formulae
given in chapter \ref{Solving}.

\section{Conclusion}

In this paper we have computed the tadpole conditions for the
orientifold projection $G + \Omega G$ of type IIB string theory on
$T^6$ for both supersymmetric and non-supersymmetric cases, where
$G$ contains a Scherk-Schwarz deformation that shifts the momenta
and gives different boundary conditions to bosons and fermions.
We have found that, when an element that breaks supersymmetry acts
in the same direction as a $Z_2$ reflection element,
anti-D5-branes must be added to cancel the tadpoles.

It is easy to generalize our results to cases for which the
breaking of supersymmetry is made by a winding shift (instead of
momentum shift). In that case, the adding of
anti-$\bar{\textrm{D}}$9-branes is necessary to cancel the
tadpoles. Anti-$\bar{\textrm{D}}$5-branes should be included when
a $Z_2$ reflection element acts perpendicular to the element that
breaks supersymmetry.
%
%
The tadpole conditions for the antibranes are similar to the ones
for the branes.

We have found complete agreement with models studied in the
literature and by adding an order two Scherk-Schwarz deformation
to the supersymmetric $T^6/Z_4$ model we was able to solve the
tadpole conditions and obtain an anomaly free massless spectrum.
We argued that the same calculations can be done for the other
inconsistent groups $Z_8$, $Z_8^{\prime}$ and $Z_{12}$.

In all the paper we have only considered D-branes sitting on one
fixed point, mainly the origin. However, it is not difficult to
study more general cases where the D5-branes are distributed on
different fixed points. We did not consider the additional freedom
of adding Wilson lines, they are non dynamical. A non-trivial
effective potential is generated for these gauge invariant
operators that dynamically breaks part of the gauge group.

This study and classification of supersymmetric and
non-supersymmetric orientifolds can be very useful for model
building \cite{Antoniadis:2000en}.
This analysis can be extended to more general orientifold groups
$G_1+\Omega G_2$, as it was discussed in the introduction, by
considering group elements which do not commute with $\Omega$ as
well as asymmetric orbifold groups \cite{Narain:1986qm}. We can
also study models with fluxes which are T-dual to orientifolds
with branes at angles. The last appear to be particularly
interesting for phenomenological applications \cite{phen1, phen2,
phen3, phen4, phen5, phen6, mlgp, Antoniadis:2004pp}.

\vskip 1cm

\centerline{\bf\Large Acknowledgments}

The authors would like to thank Elias Kiritsis, Massimo Bianchi
and Carlo Angelantonj for very useful discussions. P.
Anastasopoulos would like to thank also ``Tor Vergata" and SISSA
for hospitality.
The work of A. B. Hammou was supported by RTN contracts
HPRN-CT-2000-0131. This work was partially supported by RTN
contracts INTAS grant, 03-51-6346, RTN contracts
MRTN-CT-2004-005104 and MRTN-CT-2004-503369 and by a European
Union Excellence Grant, MEXT-CT-2003-509661.
The work of P. Anastasopoulos was supported by a Herakleitos
graduate fellowship.


\bigskip\appendix

\section{Some definitions}

Let us define some of the objects that we used in this paper. The
oscillator dependant parts are:
\bea T[^0_v]&=&\frac{1}{2\eta^2} \sum_{a,b} (-1)^{a+b+ab}
\frac{\vartheta[^a_b]}{\eta} \prod_i -2\sin \p v_i \frac{\vartheta
[^{~~a}_{b+2v_i}]}{\vartheta [^{~~1}_{1+2v_i}]}. \label{Tov}\\
T[^u_v]&=&\frac{1}{2\eta^2} \sum_{a,b} (-1)^{a+b+ab}
\frac{\vartheta[^a_b]}{\eta} \prod_i \frac{\vartheta
[^{a+2u_i}_{b+2v_i}]}{\vartheta [^{1+2u_i}_{1+2v_i}]}. \label{Tgv}
\eea
The lattice parts are:
\bea \L_{m+a,n+b} = {1 \over\eta(q) \eta(\bar{q})} \sum_{m,n}
q^{{\a'\over 4}\left({m+a\over R}+{n+b\over\a'}R\right)^2}
\bar{q}^{{\a'\over 4}\left({m+a\over R}+{n+b\over\a'}R\right)^2}
\label{LI}\eea
and the momentum and winding parts:
\bea P_{m}({i\tau_2}/2) &=& {1 \over\eta({i\tau_2}/2)} \sum_{m}
q^{{\a'\over 4}\left({m\over R}\right)^2} \label{PI}\\
W_{n}({i\tau_2}/2) &=& {1 \over\eta({i\tau_2}/2)} \sum_{n}
q^{{\a'\over 4}\left({n R \over \a'}\right)^2} \label{WI}
\eea
%
%

\section{Orientifolds}

In the appendix we will give the complete structure of the
formulae we have used in the main paper.

\subsection{Klein Bottle}

Consider an element $\a\in G$ such that $v_\a=(v_\a^1,v_\a^2,0)$.
The contribution to the Klein Bottle amplitude of this element $\a
\in G$ can be written in the form:
\bea {\cal K}_\a \sim T[^{~0}_{2v_\a}] + T[^{~g}_{2v_\a}].
\label{gf}~. \eea
In all expressions we skip the integral  $\frac{1}{2}\int
\frac{d\tau_2}{\tau_2^3}$. To extract the massless tadpole
contribution we need to perform a modular transformation $l=1/4t$
where $t$ is the loop modulus and $l$ the cylinder length and then
take the limit $l\to \infty$. After taking this limit we find:
\bea T[^{~0}_{2v_\a}] \rightarrow (1_{NS}-1_{R}) {\cal V}_3
\frac{2^3}{\prod_{l} 2~\sin 2\pi v_\a^l} \bigg(\prod_{l} 2~\cos \pi
v_\a^l\bigg)^2, \label{kv} \eea
where the factor $2^3$ comes from the modular transformation. If the orbifold
group $G$ contains a $Z_2$ factor (denoted by
$R$), it will give an extra contribution since $(\Omega R \a)^2= \a^2$:
\bea &&T[^{~~0}_{2(g_{i} v_\a)}] \rightarrow
-(1_{NS}-1_{R})\frac{1}{{\cal V}_3} \frac{2^3}{\prod_{l} 2~\sin 2\pi
v_\a^l} (2~\cos \pi v_\a^i)^2 (2~\sin \pi v_\a^j)^2~, ~~i\neq
j=1,2,
\nonumber\\
\label{kgv1}\\
&&T[^{~~0}_{2(g_3 v_a)}] \rightarrow (1_{NS}-1_{R}) {\cal V}_3
\frac{2^3}{\prod_{l} 2~\sin 2\pi v_\a^l} \bigg(\prod_{l} 2~\sin \pi
v_\a^l\bigg)^2. \label{kgv2} \eea
When we also include an order two Scherk-Schwarz deformation $h$,
we will have contributions from the twisted sector by $h$ whenever
there is a $Z_2$ element acting in the same direction. The
contribution of these sectors to the tadpoles is given by
\bea T[^{~~~~h}_{2(g_{i} v_\a)}]=T[^{~~~~h}_{2(g_{i}h v_\a)}]
\rightarrow -(1_{NS}+1_{R})\frac{2^3}{{\cal V}_3} \frac{1}{\prod_{l}
2~\sin 2\pi v_\a^l} (2~\cos \pi v_\a^i)^2 (2~\sin \pi v_\a^j)^2,
\label{kghv1} \eea
where as before $i\neq j =1,2$. There are also twisted sectors by
$R_i h$ that will produce similar results. Taking this into
account we can easily work out the equations provided in section
\ref{KleinBottle+SS}.

%
%
Putting all these together we find the amplitudes in the different
cases presented in the body of the text (\ref{k}-\ref{kkii}).

\subsection{Annulus}

To cancel the tadpoles we need to include D-branes in the spectrum
(open string sector). They are a bunch of D9-branes and in case
the group $G$ contains $g_i$-factors we also need D5$_i$-branes
extended along the $T_i^2$ torus and siting on the $R_i$-fixed
points.
The contribution of an element $\a$ can be written in the form
$\frac{1}{2}\int \frac{d\tau_2}{\tau_2^3}$:
\bea {\cal A}_\a = \bigg(Tr[\g_{\a,9}]^2 +Tr[\g_{\a,5_i}]^2\bigg)
T[^{~0}_{v_\a}] + 2 Tr[\g_{v_\a,9}] Tr[\g_{a,5_i}]
T[^{g_i}_{v_\a}] \label{av} \eea
To extract the tadpole contributions we need to perform a modular
transformation to the transverse channel $l=1/2t$ and then take
the limit $l\to \infty$ \cite{Gimon:1996rq}.
If $v_\a=(v_\a^1,v_\a^2,0)$ then, in the UV limit we find:
\bea {\cal A}_{99,\a} \rightarrow (1_{NS}-1_R){\cal V}_3
\frac{2^{-3}}{\prod_{l} 2~\sin \pi v_\a^l} Tr[\g_{\a,9}]^2. \label{av9}
\eea

If the group $G$ contains an $R$-factor than we will also have to
consider the corresponding D5-brane sectors fixed under $\a$:
\bea {\cal A}_{5_{i}5_{i},\a} \rightarrow
(1_{NS}-1_R)\frac{2^{-3}}{{\cal V}_3} \frac{1}{\prod_{l} 2~\sin \p
v_\a^l} (2\sin\p v_\a^j)^2~Tr[\g_{\a,5_{i}}]^2. \label{avi} \eea
where $i\neq j=1,2$, and
\bea {\cal A}_{5_{3}5_{3},\a} \rightarrow (1_{NS}-1_R) {\cal V}_3
\frac{2^{-3}}{\prod_{l} 2~\sin \p v_\a^l} \bigg(\prod_{l} 2\sin\p
v_\a^l\bigg)^2~Tr[\g_{\a,5_{3}}]^2. \label{avk} \eea
Including the open strings ending on different types of D-branes
we find the amplitudes (\ref{a}-\ref{aaii}).

To cancel the Klein Bottle tadpoles corresponding to the
Scherk-Schwarz deformation $h$ we need to add
$\bar{\textrm{D}}5_i$ on $R_i h$ fixed points when $R_i \in G$.
The Annulus amplitudes between two D-branes contributes
$(1_{NS}-1_R)$ whereas, the ones between a D-brane and an anti
D-brane leads $(1_{NS}+1_R)$ reflecting the breaking of
supersymmetry. The contribution to the tadpoles from the
$\bar{\textrm{D}}5_i$ branes is given by
\bea {\cal A}_{\bar{5}_{i}\bar{5}_{i},\a} \rightarrow
(1_{NS}-1_R)\frac{1}{{\cal V}_3} \frac{2^{-3}}{\prod_{l} 2~\sin \p
v_\a^l} (2\sin\p v_\a^j)^2~Tr[\g_{\a,\bar{5}_{i}}]^2. \label{avhi} \eea
%

\subsection{M\"obius Strip}

Finally, the contribution to the M\"obius strip amplitude of an
element $\a$ of $G$ can be written in the form:
\bea {\cal M}_\a & = &-\Big(Tr[\gamma^T_{\Omega
\a,9}\gamma^{-1}_{\Omega \a,9}] T[^{~0}_{v_\a}]
+Tr[\gamma^T_{\Omega R_i\a,9}\gamma^{-1}_{\Omega R_i\a,9}]
T[^{~0}_{g_iv_\a}]
\nonumber\\
&&~+Tr[\gamma^T_{\Omega \a,5_i}\gamma^{-1}_{\Omega \a,5_i}]
T[^{~0}_{g_iv_\a}] +Tr[\gamma^T_{\Omega
g_iv,5_i}\gamma^{-1}_{\Omega g_iv,5_i}] T[^{~0}_{v_\a}]\Big)~.
\nonumber \eea
where we skip again $\frac{1}{2}\int \frac{d\tau_2}{\tau_2^3}$.
The overall minus sign is conventional. We must make the same
choice of sign as for the identity element of $G$.
To extract the tadpole conditions we must perform a modular
transformation to the transverse channel by $P=T S T^2 S T$ where
$T:\t \to \t+1$ and $S:\t \to - 1/\t$ going to the $l=1/8t$.
Finally, we take the UV limit $l \to\infty$. It is not difficult
to work out the massless contributions in the different cases.

\begin{itemize}
\item For an element $v_\a=(v_\a^1,v_\a^2,0)$:
\begin{itemize}
\item[-] In case $G$ does not contain a $Z_2$ factor, the massless
contribution is:
\bea
- 2 (1_{NS}-1_R) ~{\cal V}_3~ \frac{1}{\prod_{l} 2sin2\pi v_\a^l}
Tr[\gamma^T_{\Omega \a,9}\gamma^{-1}_{\Omega \a,9}] \prod_{l}
2\cos\pi v_\a^l. \nonumber \eea
\item[-] When $G$ contains an $R$ element, we should consider also
the contribution coming from the corresponding D5-branes:
\begin{itemize}
\item[i.] If $G$ contains only a $R_3$ element:
\bea
&& -2 {\cal V}_3 \frac{(1_{NS}-1_R)}{\prod_{l} 2\sin2\pi v_\a^l}
\Bigg\{Tr[\gamma^T_{\Omega \a,9}\gamma^{-1}_{\Omega \a,9}]
\prod_{l} 2\cos\pi v_\a^l -Tr[\gamma^T_{\Omega
R_3\a,9}\gamma^{-1}_{\Omega R_3\a,9}]
\prod_{l} 2\sin\pi v_\a^l \nonumber\\
&&-\bigg(Tr[\gamma^T_{\Omega \a,5_3}\gamma^{-1}_{\Omega \a,5_3}]
\prod_{l} 2\cos\pi v_\a^l -Tr[\gamma^T_{\Omega
R_3\a,5_3}\gamma^{-1}_{\Omega R_3\a,5_3}] \prod_{l} 2\sin\pi
v_\a^l\bigg)\prod_{n} 2\sin2\pi v_\a^n\Bigg\} \nonumber \eea
\item[ii.] If only $R_{i}\in G$ for a given $i=1$ or $2$:
\bea
&& -2 \frac{(1_{NS}-1_R)}{\prod_{l} 2sin2\pi v_\a^l} \Bigg\{{\cal
V}_3 Tr[\gamma^T_{\Omega \a,9}\gamma^{-1}_{\Omega \a,9}]
\prod_{l} 2\cos\pi v_\a^l\nonumber\\
&& +Tr[\gamma^T_{\Omega R_i\a,9}\gamma^{-1}_{\Omega R_i\a,9}]
2\cos\pi v_\a^i 2\sin\pi v_\a^j +\bigg(Tr[\gamma^T_{\Omega
\a,5_i}\gamma^{-1}_{\Omega \a,5_i}]
\prod_{l} 2\cos\pi v_\a^l\nonumber\\
&& +\frac{1}{{\cal V}_3} Tr[\gamma^T_{\Omega
R_i\a,5_i}\gamma^{-1}_{\Omega R_i\a,5_i}] 2\cos\pi v_\a^i 2\sin\pi
v_\a^j\bigg) 2\sin2\pi v_\a^j\Bigg\} \nonumber \eea
\item[iii.] If all three possible $R_l\in G$ with $l=1,2,3$:
\bea
&& -2~ \frac{(1_{NS}-1_R)}{\prod_{l} 2sin2\pi v_\a^l}
\times\nonumber\\
&&\Bigg\{{\cal V}_3 \Bigg[Tr[\gamma^T_{\Omega
\a,9}\gamma^{-1}_{\Omega \a,9}] \prod_{l} 2\cos\pi v_\a^l
-Tr[\gamma^T_{\Omega R_3\a,9}\gamma^{-1}_{\Omega R_3\a,9}]
\prod_{l} 2\sin\pi v_\a^l \nonumber\\
&&-\bigg(Tr[\gamma^T_{\Omega \a,5_3}\gamma^{-1}_{\Omega \a,5_3}]
\prod_{l} 2\cos\pi v_\a^l -Tr[\gamma^T_{\Omega
R_3\a,5_3}\gamma^{-1}_{\Omega R_3\a,5_3}] \prod_{l} 2\sin\pi
v_\a^l\bigg)\prod_{n} 2\sin2\pi v_\a^n\Bigg]
\nonumber\\
&&+\frac{1}{{\cal V}_3} \sum_{i\neq j=1,2}
\bigg(Tr[\gamma^T_{\Omega R_i\a,5_i}\gamma^{-1}_{\Omega
R_i\a,5_i}]
2\cos\pi v_\a^i 2\sin\pi v_\a^j\nonumber\\
&&- Tr[\gamma^T_{\Omega R_j\a,5_i}\gamma^{-1}_{\Omega R_j\a,5_i}]
2\sin\pi v_\a^i 2\cos\pi v_\a^j\bigg)2\sin2\pi v_\a^j
\nonumber\\
&& +\sum_{i\neq j=1,2} Tr[\gamma^T_{\Omega
R_i\a,9}\gamma^{-1}_{\Omega R_i\a,9}] 2\cos\pi v_\a^i 2\sin\pi
v_\a^j
\nonumber\\
&& +\sum_{i\neq j=1,2} Tr[\gamma^T_{\Omega
R_i\a,5_3}\gamma^{-1}_{\Omega R_i\a,5_3}]
2\cos\pi v_\a^i 2\sin\pi v_\a^j \prod_{l} 2\sin2\pi v_\a^l\nonumber\\
&& +\sum_{i\neq j=1,2}Tr[\gamma^T_{\Omega
\a,5_i}\gamma^{-1}_{\Omega \a,5_i}] \prod_{l} 2\cos\pi v_\a^l
2\sin2\pi v_\a^j
\nonumber\\
&& +\sum_{i\neq j=1,2} Tr[\gamma^T_{\Omega
R_3\a,5_i}\gamma^{-1}_{\Omega R_3\a,5_i}] 2\sin\pi v_\a^i 2\sin\pi
v_\a^j 2\sin2\pi v_\a^j\Bigg\}. \nonumber \eea
\end{itemize}
\end{itemize}
\item For an element $v_\a=(v_\a^1,v_\a^2,v_\a^3)$:
\begin{itemize}
\item[-] When $G$ contains no $R$ factors we have:
\bea
- 2 (1_{NS}-1_R) \frac{1}{\prod_{l} 2\sin2\pi v_\a^l}
Tr[\gamma^T_{\Omega \a,9}\gamma^{-1}_{\Omega \a,9}] \prod_{l}
2\cos\pi v_\a^l. \nonumber \eea
\item[-] If $G$ contains $R$ factors:
\begin{itemize}
\item[i.] If there is only one $R_i \in G$ for a given $i$:
\bea
&& - 2 \frac{(1_{NS}-1_R)}{\prod_{l} 2\sin2\pi v_\a^l}
\Bigg\{Tr[\gamma^T_{\Omega \a,9}\gamma^{-1}_{\Omega \a,9}]
\prod_{l} 2\cos\pi v_l\nonumber\\
&& - Tr[\gamma^T_{\Omega R_i\a,9}\gamma^{-1}_{\Omega R_i\a,9}]
2\cos\pi v_\a^i \prod_{l\neq i} 2\sin\pi v_\a^l
-\bigg(Tr[\gamma^T_{\Omega \a,5_i}\gamma^{-1}_{\Omega \a,5_i}]
\prod_{l} 2\cos\pi v_\a^l\nonumber\\
&& - Tr[\gamma^T_{\Omega R_i\a,5_i}\gamma^{-1}_{\Omega R_i\a,5_i}]
2\cos\pi v_\a^i \prod_{l\neq i} 2\sin\pi v_\a^l\bigg) \prod_{k\neq
i}2\sin2\pi v_\a^k\Bigg\} \nonumber \eea
\item[ii.] If all possible $R_l\in G$ with $l=1,2,3$:
\bea
&& -2 \frac{(1_{NS}-1_R)}{\prod_{l}
2\sin2\pi v_\a^l}\times\nonumber\\
&&~~~\Bigg\{Tr[\gamma^T_{\Omega \a,9}\gamma^{-1}_{\Omega \a,9}]
\prod_{l} 2\cos\pi v_\a^l -\sum_{i=1}^3 Tr[\gamma^T_{\Omega
R_i\a,9}\gamma^{-1}_{\Omega R_i\a,9}]
2\cos\pi v_\a^i \prod_{l\neq i} 2\sin\pi v_\a^l\nonumber\\
&&~~~-\sum_{i=1}^3 \bigg(Tr[\gamma^T_{\Omega
\a,5_i}\gamma^{-1}_{\Omega \a,5_i}] \prod_{l} 2\cos\pi v_\a^l
\nonumber\\
&&~~~-Tr[\gamma^T_{\Omega R_i\a,5_i}\gamma^{-1}_{\Omega
R_i\a,5_i}] 2\cos\pi v_\a^i \prod_{l\neq i} 2\sin\pi v_\a^l\bigg)
\prod_{k\neq i} 2\sin2\pi v_\a^k\Bigg\} \nonumber \eea
\end{itemize}
\end{itemize}
\end{itemize}

Requiring that the M\"obius strip transverse channel amplitude to
be the mean value between the transverse channel Klein Bottle and
Annulus amplitudes gives us the constraints (\ref{consta}) on the
matrices $\g_{\a,I}$ and $\g_{\Omega.\a,I}$.

\end{document}